\documentclass[journal]{IEEEtran}
\usepackage{array}
\usepackage[caption=false,font=normalsize,labelfont=sf,textfont=sf]{subfig}
\usepackage{textcomp}
\usepackage{stfloats}
\usepackage{url}
\usepackage{verbatim}
\usepackage{graphicx}
\usepackage{cite}
\usepackage{hyperref}
\usepackage[T1]{fontenc}
\usepackage{fix-cm}
\usepackage{float}
\usepackage{tabularx}
\usepackage{booktabs}
\usepackage{threeparttable}
\usepackage{adjustbox}
\usepackage{algorithm}
\usepackage{algorithmic}
\usepackage{threeparttable}
\usepackage{booktabs}
\usepackage{multirow}
\usepackage{makecell}
\usepackage{tikz}
\usepackage{textcomp}
\usepackage{amsmath,amssymb}
\newcolumntype{C}{>{\centering\arraybackslash}X}
\newcolumntype{L}{>{\raggedright\arraybackslash}X}
\begin{document}

\title{HEDN: A Hard-Easy Dual Network with Source Reliability Assessment for Cross-Subject EEG Emotion Recognition}

\author{\IEEEauthorblockN{Qiang Wang, Liying Yang, Jiayun Song, Yifan Bai, Jingtao Du}

\thanks{This work was supported by the National Natural Science Foundation of China under Grants 62374121 and 61974109. (\emph{Corresponding author : Liying Yang.})}
\thanks{Qiang Wang, Liying Yang, Jiayun Song, Yifan Bai and Jingtao Du are with the School of Computer Science and Technology, Xidian University, Xi'an 710071, China (e-mail: qiang.wang@stu.xidian.edu.cn; yangliying1208@163.com; sjy@stu.xidian.edu.cn; yifan\_bai@stu.xidian.edu.cn;  jtdu@stu.xidian.edu.cn).}
\thanks{The source code is available at \emph{https://github.com/qwangwl/HEDN}}
}

\maketitle

\begin{tikzpicture}[remember picture, overlay]
\node[anchor=south, yshift=10pt] at (current page.south) {
    \parbox{\textwidth}{\centering \footnotesize \textit{This work has been submitted to the IEEE for possible publication. Copyright may be transferred without notice, after which this version may no longer be accessible.}}
};
\end{tikzpicture}

\begin{abstract}
Cross-subject electroencephalography (EEG) emotion recognition remains a major challenge in brain-computer interfaces (BCIs) due to substantial inter-subject variability. Multi-Source Domain Adaptation (MSDA) offers a potential solution, but existing MSDA frameworks typically assume equal source quality, leading to negative transfer from low-reliability domains and prohibitive computational overhead due to multi-branch model designs. To address these limitations, we propose the Hard-Easy Dual Network (HEDN), a lightweight reliability-aware MSDA framework. HEDN introduces a novel Source Reliability Assessment (SRA) mechanism that dynamically evaluates the structural integrity of each source domain during training. Based on this assessment, sources are routed to two specialized branches: an Easy Network that exploits high-quality sources to construct fine-grained, structure-aware prototypes for reliable pseudo-label generation, and a Hard Network that utilizes adversarial training to refine and align low-quality sources. Furthermore, a cross-network consistency loss aligns predictions between branches to preserve semantic coherence. Extensive experiments conducted on SEED, SEED-IV, and DEAP datasets demonstrate that HEDN achieves state-of-the-art performance across both cross-subject and cross-dataset evaluation protocols while reducing adaptation complexity.
\end{abstract}


\section{Introduction}
\IEEEPARstart{E}{motion} plays a critical role in human cognition, influencing perception, decision-making, and social interaction. The capability to automatically recognize emotional states from neural activity holds significant potential for next-generation human–computer interaction (HCI), facilitating applications such as empathetic virtual agents, adaptive educational systems, and the early diagnosis of neuropsychiatric disorders~\cite{ali2016eeg, brunner2015bnci}. Among various physiological modalities, electroencephalography (EEG) provides a direct, non-invasive, and temporally resolved representation of the brain’s affective and cognitive responses~\cite{li2022eeg}. These properties position EEG emotion recognition as a highly promising research direction.

However, individual differences in neuroanatomy and emotional expression introduce substantial distributional discrepancies in EEG recordings collected from different subjects~\cite{wang2021deep, wang2024dmmr, shen2022contrastive}. This inter-subject variability limits model generalization and poses significant challenges for real-world deployment. Domain adaptation (DA) has therefore become a core research direction in cross-subject EEG emotion recognition, aiming to reduce distribution mismatches between labeled source domains and unlabeled target domains to improve generalization~\cite{zhong2020eeg}. Multi-source domain adaptation (MSDA) extends this paradigm by treating each subject as an independent domain~\cite{zhao2020multi}, thereby avoiding the unrealistic assumption of a single unified source distribution~\cite{li2019multisource, chen2021ms}.

Despite its progress, two major limitations in existing MSDA research have increasingly surfaced. First, most approaches assign equal importance to all source domains, failing to account for differences in signal quality, intra-class consistency, and domain reliability~\cite{chen2024gddn}. Low-quality sources may introduce noise and ambiguous class boundaries, ultimately leading to negative transfer~\cite{ma2023cross}. Second, maintaining independent adaptation branches for each source domain substantially increases computational and memory overhead, making such architectures impractical for real-time brain–computer interfaces (BCIs).

\begin{figure}[!t]
\centering
\includegraphics[width=0.8\linewidth]{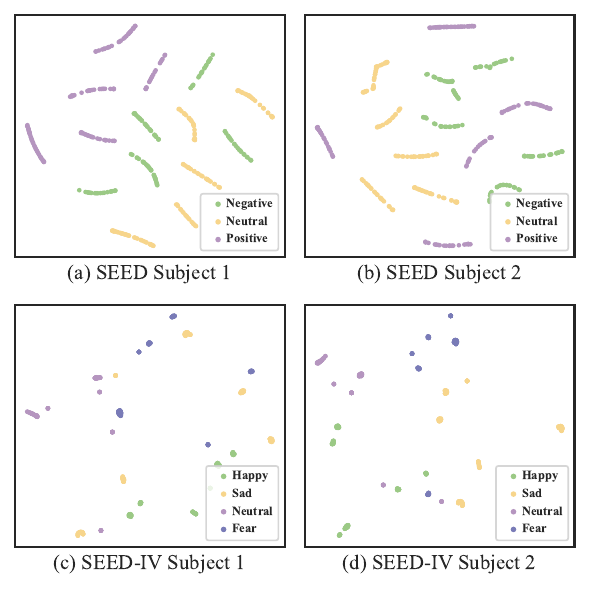}
\caption{t-SNE visualizations illustrating intra-subject heterogeneity in EEG emotion recognition. Distinct multi-cluster structures within each emotional state are evident in (a–b) SEED subjects and (c–d) SEED-IV subjects, highlighting substantial intra-class variability.}
\label{fig:intra-subject-heterogeneity}
\end{figure}

Recent studies show that EEG emotion signals display substantial intra-subject heterogeneity, where samples from the same emotional state can vary markedly within a single individual~\cite{wang2025fine}, as illustrated in Fig.~\ref{fig:intra-subject-heterogeneity}. The t-SNE visualizations~\cite{maaten2008visualizing} reveal that EEG samples belonging to the same emotional category often form multiple distinct clusters, even within a single subject. These latent cluster patterns reflect individualized neural responses to emotional stimuli and are shaped by contextual factors such as stress, attentional state, and task engagement~\cite{ahn2016exploring, mehmood2023deep}. Specifically, in EEG-based affective computing datasets, this phenomenon exhibits strong trial dependency: samples within the same cluster frequently originate from the same stimulus trial because different video clips not only trigger distinct affective responses but also vary in cognitive load, attention modulation, and emotional intensity. Importantly, recent work suggests that these reproducible latent substructures can serve as a strong prior for fine-grained prototype construction, thereby enabling more reliable pseudo-labeling for target-domain samples~\cite{ji2020improved, wang2025fine}.

This observation offers a promising solution to the reliability and efficiency bottlenecks of existing MSDA frameworks. We propose a key hypothesis: effective adaptation does not require a dedicated branch for every source domain. Instead, a small number of specialized branches, assigned adaptively based on domain reliability, are sufficient. We argue that the structural quality of a source determines its functional role in adaptation. Specifically, high-reliability sources, characterized by well-formed latent cluster structures, are ideal for constructing fine-grained prototypes~\cite{snell2017prototypical} that provide structural guidance and generate high-confidence pseudo-labels for the target domain. In contrast, low-reliability sources, characterized by ambiguous class boundaries, are processed in a refinement branch designed to stabilize their embeddings before contributing supervisory signals. By dynamically routing sources to these distinct processing pathways, our design simultaneously ensures precision in knowledge transfer and significant gains in architectural efficiency.

To this end, we propose the Hard-Easy Dual Network (HEDN), a lightweight reliability-aware MSDA framework that dynamically differentiates source domains based on their structural quality. HEDN integrates a Source Reliability Assessment (SRA) mechanism that dynamically evaluates the structural reliability of each source domain during training. Reliability is quantified via classification stability, reflecting a domain’s suitability for prototype construction. The source with the highest structural quality is designated the Easy Source and routed to the Easy Network, while the lowest-quality source is labeled the Hard Source and directed to the Hard Network. The Easy Network leverages the well-organized multi-cluster structure of the Easy Source to build domain-specific prototypes and generate high-confidence pseudo-labels for the target domain through structure-aware prototype matching. To further enhance structural fidelity, it employs cluster-level contrastive learning in a dedicated embedding space, explicitly promoting intra-cluster compactness and inter-cluster separability—thus refining prototype representations and improving pseudo-label accuracy. Conversely, the Hard Network enhances the discriminability of the Hard Source through adversarial training, improving robustness against noise, complex distributional shifts, and intra-class ambiguity. Additionally, a cross-network consistency loss aligns the two branches by constraining the Hard Network’s predictions to match the high-confidence outputs of the Easy Network, preserving semantic coherence. Through this design, HEDN transforms structural variability into an adaptive learning signal: reliable sources serve as anchors, while unreliable ones are progressively refined—enabling scalable and robust cross-subject EEG emotion recognition.

The main contributions of this paper are summarized as follows:

\begin{itemize}
\item We propose HEDN, a reliability-aware MSDA framework that dynamically routes the most and least reliable source domains to two specialized branches, replacing per-source adaptation with a lightweight, scalable architecture.
\item We exploit intra-subject heterogeneity as a structural prior: the Easy Network constructs fine-grained prototypes and produces high-confidence pseudo-labels, which are further refined via cluster-level contrastive learning to enhance structural consistency.
\item Extensive experiments on SEED, SEED-IV, and DEAP demonstrate that HEDN consistently outperforms state-of-the-art methods in accuracy and robustness while maintaining high computational efficiency.
\end{itemize}

\section{Related Work}

\subsection{Multi-Source Domain Adaptation}

Multi-source domain adaptation aims to leverage multiple labeled source domains to improve generalization on an unlabeled target domain, a setting that naturally arises in EEG-based affective computing due to substantial inter-subject variability. Unlike single-source adaptation, MSDA must address not only global distribution shifts but also the heterogeneous quality and transferability of different source subjects. Therefore, an effective MSDA framework must balance two objectives: (i) aligning representations across multiple domains and (ii) selectively exploiting source domains according to their intrinsic reliability and relevance to the target.

A substantial body of MSDA research assumes uniform treatment of sources, implicitly presuming that all domains contribute equally to adaptation. Representative methods such as MS-MDA~\cite{chen2021ms} disentangle domain-invariant and domain-specific components through a shared–private architecture, while S$^{2}$A$^{2}$-MSDA~\cite{yang2024spectral} introduces spectral–spatial attention and pairwise alignment modules to refine cross-domain consistency. Although these approaches outperform single-source baselines, their source-equality assumption overlooks the fact that some subjects may be poorly aligned with the target, resulting in negative transfer.

To alleviate this issue, subsequent studies have incorporated source selection or weighting mechanisms. MSGDAN~\cite{wang2024multi} measures maximum mean discrepancy (MMD) to retain only the most similar sources, whereas LSA-MMFT~\cite{she2022multi} evaluates transferability via label-similarity consistency. While these strategies effectively filter out harmful sources, they rely on static decisions that may discard potentially valuable information and do not differentiate processing according to source quality during training.

In contrast to conventional MSDA methods, HEDN retains all source domains and dynamically evaluates their reliability via a lightweight SRA mechanism throughout the training process. Sources are routed into the Easy and Hard networks based on their quality: the Easy Network exploits high-quality sources for prototype construction and reliable pseudo-labels, while the Hard Network progressively enhances low-quality sources. This difficulty-aware design maximizes the use of informative sources and mitigates interference from noisier domains, offering a more flexible and robust alternative to static MSDA approaches.

\subsection{Prototype Learning}

Prototype learning provides a principled approach to modeling semantic structure by constructing representative centroid vectors that summarize subdomain-level distributions. In cross-subject EEG emotion recognition, this paradigm has served as a cornerstone for mitigating noise and label inconsistency. By abstracting complex neural signals into compact representations, prototype-based models facilitate structural alignment and enhance the reliability of pseudo-labeling across heterogeneous subjects.

Existing studies predominantly employ global, class-level prototypes to capture emotion-specific structure. PR-PL~\cite{zhou2023pr} forms emotion prototypes to improve robustness against label noise, while MSPR-GNN~\cite{guo2023multi} extends prototype learning to MSDA through a graph-based module that aligns subject-specific prototypes. ADANN~\cite{hong2024adaptive} further integrates prototype-based contrastive learning with conditional distribution alignment to enhance generalization across subjects. Although effective, these methods assume relative within-class homogeneity and thus overlook the fine-grained, multi-cluster structures commonly present in EEG signals.

To utilize this intra-class diversity, recent works such as DBPM~\cite{wang2025fine} adopt density-based clustering to uncover multi-cluster substructures and construct fine-grained prototypes that better capture subject-specific neural patterns, improving the accuracy of pseudo-label propagation. However, DBPM evaluates all source domains equally and lacks mechanisms to distinguish structurally reliable sources from those dominated by ambiguous clusters. As a result, prototype matching may degrade when low-quality sources exert disproportionate influence.

In contrast, HEDN combines fine-grained prototype modeling with reliability-guided source selection. HEDN identifies sources with compact, stable structures and routes them to the Easy Network for cluster-aware prototype construction and contrastive learning. By strictly limiting prototype modeling to reliable domains, HEDN ensures the generation of high-confidence pseudo-labels while preventing interference from low-quality data, effectively advancing the robustness of prototype learning in cross-subject scenarios.

\section{Methods}

\begin{figure*}[!t]
  \centering
  \includegraphics[width=\textwidth]{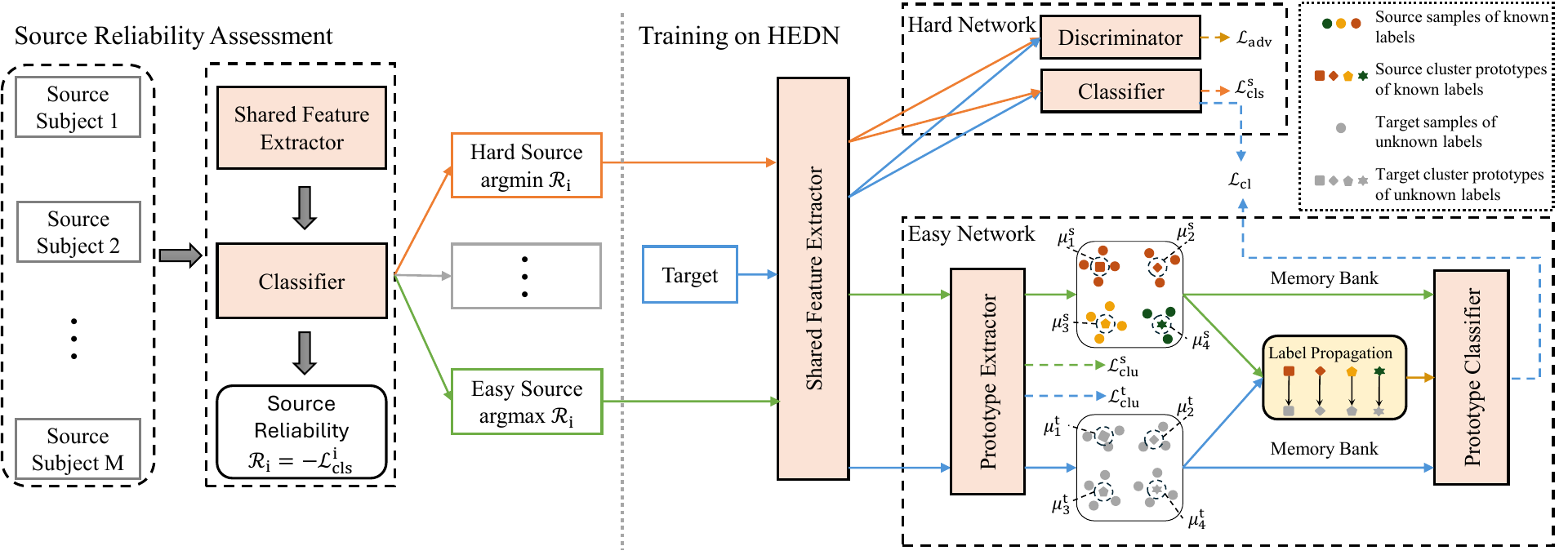}
  \caption{Overview of the proposed HEDN. The framework dynamically routes each source domain to either the Hard Network or Easy Network based on its assessed reliability. The Hard Network employs adversarial training to enhance discriminability of low-reliability sources, while the Easy Network leverages structure-aware prototype learning to exploit multi-cluster patterns in high-reliability sources for stable pseudo-label generation. Both branches share a common feature extractor and are jointly optimized via cross-network consistency constraints.}
  \label{fig:framework}
\end{figure*}

In multi-source cross-subject EEG emotion recognition, we are given $M$ labeled source domains $\mathcal{S} = \{\mathcal{D}_s^i\}_{i=1}^M$, where each source domain $\mathcal{D}_s^i = \{(x_j^i, y_j^i)\}_{j=1}^{N_i}$ contains $N_i$ EEG samples from the $i$-th subject with corresponding emotion labels. The target domain $\mathcal{D}_t = \{x_j^t\}_{j=1}^{N_t}$ consists of $N_t$ unlabeled EEG samples from a new subject. The objective is to leverage knowledge from the multiple source domains to train a model that generalizes robustly to the target domain.

Figure~\ref{fig:framework} presents the overall architecture of the proposed HEDN. The framework consists of three key components: (1) a lightweight SRA module that dynamically evaluates the structural reliability of each source domain; (2) a Hard Network that enhances transferability of low-quality sources through adversarial training; and (3) an Easy Network that exploits high-quality sources via structure-aware prototype learning for stable knowledge transfer. A shared feature extractor $f(\cdot)$ generates common representations for both branches. By dynamically routing source domains to the appropriate network branch based on their estimated difficulty, HEDN enables targeted adaptation strategies that enhance overall transfer effectiveness.

\subsection{Source Reliability Assessment}

The Source Reliability Assessment (SRA) module is designed to dynamically determine, at each training iteration, which source domain provides the most reliable supervision for prototype construction and pseudo-label generation (Easy Source) and which source is the least reliable and requires refinement (Hard Source).

To implement this efficiently, we use the classification cross-entropy loss as a lightweight yet effective proxy for structural reliability. A lower loss indicates compact intra-class distributions and consistent labels, reflecting higher structural quality of the source domain, whereas a higher loss corresponds to ambiguous or noisy structures~\cite{han2018co, hu2023swl}. Crucially, this metric requires no auxiliary models or domain-level statistics; it is computed intrinsically during training and can be evaluated in a single forward pass, ensuring minimal computational overhead.

Formally, at each training iteration, a mini-batch is sampled from every source domain and evaluated using the classifier $h(\cdot)$ associated with the Hard Network. Given the shared feature extractor $f(\cdot)$, the cross-entropy loss for the $i$-th source domain is computed as:
\begin{equation}
    \label{eq:cls_loss}
    \mathcal{L}_{\text{cls}}^{i} = -\frac{1}{B}\sum_{j=1}^{B} y_j^i \log(h(f(x_j^i))),
\end{equation}
where $B$ is the mini-batch size and $y_j^i$ is the one-hot label of sample $x_j^i$ from the $i$-th source domain. The reliability score is then defined as the negative loss:
\begin{equation}
    \label{eq:R}
    \mathcal{R}_i = -\mathcal{L}_{\text{cls}}^{i},
\end{equation}
such that higher \(\mathcal{R}_i\) corresponds to greater structural reliability. 

Based on these scores, we assign roles to the source domains in the current iteration: the domain with the highest score is selected as the Easy Source, and the one with the lowest score is designated the Hard Source. All other domains are ignored for that iteration:
\begin{equation} 
    \label{eq:select} 
    k_{\text{easy}} = \arg\max_i \mathcal{R}_i, \quad k_{\text{hard}} = \arg\min_i \mathcal{R}_i. 
\end{equation}

The mini-batch from the Easy Source $\mathcal{B}_{k_{\text{easy}}}$ is routed to the Easy Network for constructing fine-grained prototypes and generating structure-aware pseudo-labels. The Hard Source batch $\mathcal{B}_{k_{\text{hard}}}$ is sent to the Hard Network, where adversarial training enhances robustness against noise and distributional discrepancies. This dynamic, reliability-based routing ensures that structurally informative sources drive prototype modeling while low-quality sources are selectively strengthened, effectively mitigating negative transfer.

Notably, SRA operates exclusively at the mini-batch level and requires only a single forward evaluation through the classifier without computing domain-level statistics. This design ensures computational efficiency and scalability in large-scale multi-source settings. Furthermore, by selectively processing only the most representative mini-batches from each source rather than exhaustively traversing all source-domain samples, SRA reduces redundant computation and accelerates convergence.

\subsection{Hard Network}

The Hard Network is designed to process source domains identified by the SRA module as having low structural reliability. Notably, SRA evaluates each source solely based on its internal structural coherence, quantified through instantaneous classification loss, rather than its similarity to the target domain. As a result, Hard Sources may still contain meaningful labeled information relevant to the target domain and therefore should not be discarded. Instead, the Hard Network aims to refine and stabilize these structurally weak domains, enabling them to contribute effectively to cross-subject knowledge transfer. Moreover, to further address distributional discrepancies inherent in cross-subject EEG emotion recognition, we integrate a Domain-Adversarial Neural Network (DANN)~\cite{ganin2016domain} to align feature distributions across source and target domains.

The Hard Network architecture comprises three components: a shared feature extractor $f(\cdot)$, an emotion classifier $h(\cdot)$, and a domain discriminator $d(\cdot)$. 
Here, $f(\cdot)$ encodes raw EEG signals into latent representations, $h(\cdot)$ encourages emotion-discriminative feature learning, and $d(\cdot)$ differentiates source and target features through adversarial training, thereby guiding $f(\cdot)$ to produce domain-invariant representations. To enable end-to-end optimization, a Gradient Reversal Layer (GRL) is employed to reconcile the conflicting objectives of feature extraction and adversarial adaptation.

Formally, at each iteration, the Hard Network optimizes the following objective over the Hard Source mini-batch and a target-domain mini-batch:
\begin{equation}
\label{eq:loss_hard}
\mathcal{L}_{\textrm{hard}} = \mathcal{L}_{\textrm{cls}}^s + \lambda_1 \mathcal{L}_{\textrm{adv}},
\end{equation}
where $\lambda_1$ is a hyperparameter that balances the contribution of the classification and adversarial losses, following the exponential growth schedule used in existing studies~\cite{liu2024capsnet,zhou2023pr}. 

The classification loss for the Hard Source is defined as:
\begin{equation}
\mathcal{L}_{\textrm{cls}}^s = -\frac{1}{B} \sum_{i=1}^{B} y_i^s \log\left(h(f(x_i^s))\right),
\end{equation}
and the adversarial loss for domain alignment between the Hard Source and the target domain is defined as:
\begin{equation}
\begin{aligned}
\mathcal{L}_{\text{adv}} = & -\frac{1}{B} \sum_{i=1}^{B} \log d(f(x_i^s)) \\
& - \frac{1}{B} \sum_{i=1}^{B} \log (1 - d(f(x_i^t))).
\end{aligned}
\end{equation}

By jointly minimizing the classification loss and the adversarial loss, the Hard Network fulfills two key objectives: (1) it enhances the discriminability of Hard Source features by reducing classification errors, and (2) it alleviates cross-domain distribution mismatch. This dual objective effectively improves the utility of Hard Sources, enabling them to participate meaningfully in scalable cross-subject EEG emotion recognition.

\subsection{Easy Network}

The Easy Network processes source domains that the SRA module identifies as structurally reliable. These domains exhibit low internal classification loss, compact intra-class distributions, and stable labels. Because of this structural consistency, they provide high-quality supervision suitable for prototype refinement and structure-aware cross-domain adaptation. 

To leverage their reliability, the Easy Network adopts a prototype-based learning paradigm. The architecture includes a shared feature extractor $f(\cdot)$, a prototype feature extractor $g(\cdot)$, and a prototype-driven classifier $\psi$. While $f(\cdot)$ captures shared EEG representations, $g(\cdot)$ emphasizes discriminative structure that supports prototype construction.  Its primary objective is to transfer class-conditional knowledge to the target domain through domain-specific prototype construction and cluster-level contrastive learning, thereby generating high-confidence pseudo-labels that preserve semantic and structural fidelity.

\subsubsection{Domain-Specific Prototype Management}

To preserve subject-specific structural properties, an independent prototype memory bank is maintained for each domain:
\begin{equation}
    \label{eq:source_memory}
    \mathbb{M} = \{\mathcal{U}_s^1, \dots, \mathcal{U}_s^M, \mathcal{U}_t\},
\end{equation}
where $\mathcal{U}_s^i$ contains cluster-wise prototypes for $i$-th source domain, and $\mathcal{U}_t$ is the prototype bank for the target domain.

\textbf{Initialization via Pre-training Clustering.}  
Before training, all samples from each source domain and the target domain are clustered independently using DBSCAN~\cite{schubert2017dbscan}. While intra-class clusters in EEG datasets often align with trial structure, trial labels are not used as cluster indicators because they may be missing or noisy and reflect stimulus presentation rather than intrinsic neural geometry~\cite{wang2025fine}. In contrast, DBSCAN has been shown to provide more robust cluster partitions by adapting to local density variations and filtering out outliers~\cite{abdolzadegan2020robust}. The resulting cluster assignments ($\mathcal{C}_s^i$ for source $i$ and $\mathcal{C}_t$ for the target) are fixed and retained throughout training to provide consistent structural supervision. Initial prototypes are computed as the centroids of these pre-defined clusters.

\textbf{Batch-wise Prototype Update.}
During training, when SRA routes a mini-batch from the $k$-th source to the Easy Network, only the corresponding memory banks $\mathcal{U}_s^k$ and $\mathcal{U}_t$ are updated. For each cluster $j$ present in the batch, the instantaneous prototype is computed as:
\begin{equation}
    \label{eq:com_proto_easy}
    \hat{\mu}_{k,j} = \frac{1}{|\mathcal{B}_{k,j}|} \sum_{x \in \mathcal{B}_{k,j}} g(f(x)),
\end{equation}
where $\mathcal{B}_{k,j}$ is the set of samples in the current batch belonging to cluster $j$ of source $k$.

To ensure temporal stability while accommodating evolving supervision quality, prototypes are updated using an Exponential Moving Average (EMA):
\begin{equation}
    \label{eq:ema_source}
    \mu_{k,j} = \gamma_s \mu_{k,j} + (1 - \gamma_s) \hat{\mu}_{k,j}.
\end{equation}

Similarly, when target samples are processed, we update the target cluster prototypes $\mu_{t,i} \in \mathcal{U}_t$ using a separate momentum $\gamma_t$:
\begin{equation}
\label{eq:ema_target}
\mu_{t,i} = \gamma_t \mu_{t,i} + (1 - \gamma_t) \hat{\mu}_{t,i}.
\end{equation}
Here, $\gamma_s, \gamma_t \in [0,1)$ are hyperparameters. Typically, $\gamma_t$ is set differently from $\gamma_s$ to account for the disparity in update frequency: source domains participate intermittently depending on SRA routing, whereas the target domain is continuously involved throughout training.

\subsubsection{Prototype-Guided Label Propagation}

Inspired by prior work~\cite{wang2025fine}, these robust prototypes are further leveraged to generate high-quality pseudo-labels for the target domain. For each target cluster prototype $\mu_{t,i} \in \mathcal{U}_t$, we compute cosine similarity to all prototypes in $\mathcal{U}_s^k$:
\begin{equation}
    \label{eq:cos_sim}
    \text{sim}(\mu_{t,i}, \mu_{k,j}) = \frac{\mu_{t,i} \cdot \mu_{k,j}}{\|\mu_{t,i}\|_2 \|\mu_{k,j}\|_2}.
\end{equation}
The label of the most similar source prototype is assigned to the entire target cluster, defining prototype-driven classifier $\psi$:
\begin{equation}
    \label{eq:label_map}
\psi : \mathcal{X}_t \rightarrow \mathcal{C}_t \rightarrow \mathcal{Y}_s,
\end{equation}
which bridges unlabeled target samples and semantic emotion categories through two sequential steps. First, each target sample is assigned to a structural cluster ($\mathcal{X}_t \rightarrow \mathcal{C}_t$) based on its feature proximity to cluster prototypes in $\mathcal{U}_t$. Second, the entire cluster inherits the emotion category of its matched source prototype ($\mathcal{C}_t \rightarrow \mathcal{Y}_s$).

This structured propagation enables the Easy Network to leverage unlabeled target data as supervised-like signals, producing high-confidence, structure-aware pseudo-labels that preserve semantic and cluster-level fidelity.

\subsubsection{Cluster Contrastive Learning}

To enforce structural consistency within the precomputed density-based clusters and refine the semantic representations learned by the Easy Network, we adopt a cluster-level supervised contrastive loss~\cite{khosla2020supervised}. Specifically, we employ a lightweight bottleneck MLP network as the prototype-specific feature extractor $g(\cdot)$, which maps the encoder output $f(x)$ into a low-dimensional embedding space where cluster structures can be more explicitly captured. This bottleneck reduces task-irrelevant noise and encourages the model to retain only cluster-consistent information, facilitating stable cross-domain alignment.

Given a mini-batch comprising samples from the SRA-selected Easy Source domain and the target domain, the supervised contrastive loss is computed using the fixed cluster assignments from the initial clustering. For each anchor sample $i$, let $\mathcal{P}(i)$ denote the set of other samples in the batch that belong to the same precomputed cluster. The cluster-level contrastive loss is formulated as:
\begin{equation}
\label{eq:clu_loss_easy}
\mathcal{L}_{\text{clu}} = -\frac{1}{B} \sum_{i=1}^{B} \frac{1}{|\mathcal{P}(i)|} \sum_{p \in \mathcal{P}(i)} \log \frac{\exp({s_{ip}/\tau})}{\sum_{j \neq i} \exp({s_{ij}/\tau})},
\end{equation}
where $s_{ij} = \cos(\mathbf{z}_i, \mathbf{z}_j)$ is the cosine similarity between projected semantic features $\mathbf{z} = g(f(x))$, and $\tau$ is a temperature hyperparameter that controls the concentration of the similarity distribution.

By minimizing $\mathcal{L}_{\text{clu}}$, the network pulls together samples from the same cluster and pushes apart those from different clusters in the embedding space produced by $g(\cdot)$. This optimization increases intra-cluster compactness and inter-cluster separability, enhancing prototype quality. Consequently, cluster centroids become more representative, improving the reliability of prototype-based pseudo-label propagation for the target domain.

\subsection{Cross-Network Consistency Learning}

The Hard and Easy Networks are specialized to handle source domains of differing complexity, yet they share the common objective of accurately classifying the target domain. A potential challenge in this dual-stream architecture is semantic divergence: because the two networks are optimized on disjoint source subsets (Hard vs. Easy) with distinct objectives (adversarial alignment vs. prototype-guided label propagation), their predictions for the same target sample may conflict. This is particularly relevant for the Hard Network, whose adversarial training, while effective at aligning marginal distributions, can sometimes blur fine-grained class boundaries due to the inherent instability of domain discriminator optimization of the target domain.

To mitigate this issue and foster synergistic adaptation, we introduce a Cross-Network Consistency Loss that aligns the output of the Hard Network with the high-confidence pseudo-labels generated by the Easy Network. The rationale is that the Easy Network, trained on structurally reliable sources with explicit clustering constraints, provides more reliable and structurally consistent supervision for the target domain.

Specifically, for a given target mini-batch, let $\hat{y}_{t}^e$ denote the high-confidence pseudo-label generated by the Easy Network via the prototype-driven classifier $\psi$ (as defined in Eq.~\ref{eq:label_map}). Let $p_{t}^h = h(f(x_{t}))$ be the softmax prediction of the Hard Network’s emotion classifier for the same sample. The consistency loss is defined as the cross-entropy between these two distributions:
\begin{equation}
    \label{eq:consistency_loss_refined}
    \mathcal{L}_{\text{cl}} = -\frac{1}{B} \sum_{i=1}^{B} \hat{y}_{t,i}^e \log(p_{t,i}^h).
\end{equation}

By minimizing $\mathcal{L}_{\text{cl}}$, we achieve two critical effects. First, it acts as structural regularization: by distilling the fine-grained cluster structure learned by the Easy Network into the Hard Network, it counteracts the boundary-blurring effect of adversarial training and preserves discriminative class semantics. Second, it enables collaborative adaptation: knowledge distilled from high-quality sources is actively propagated to guide the interpretation of low-quality sources, establishing a mutually reinforcing feedback loop that enhances overall transfer robustness and mitigates negative transfer.

\subsection{Step-wise Joint Optimization}

\begin{algorithm}[t]
\caption{Training Algorithm of HEDN}
\label{alg:training}
\begin{algorithmic}[1]
\REQUIRE Source domains $\mathcal{S} = \{\mathcal{D}_s^i\}_{i=1}^{M}$, target domain $\mathcal{D}_t$
\ENSURE Trained HEDN model
\STATE \textbf{Initialization:} Perform DBSCAN on each $\mathcal{D}_s^i$ and $\mathcal{D}_t$ to obtain fixed cluster assignments.
\STATE Initialize source and target prototypes using cluster centroids; store in memory banks $\{\mathcal{U}_s^i\}_{i=1}^M$ and $\mathcal{U}_t$.
\FOR{each training iteration}
    \STATE Sample a mini-batch $\mathcal{B}_t$ from $\mathcal{D}_t$ and $\{\mathcal{B}_s^i\}_{i=1}^M$ from source domains $\{\mathcal{D}_s^i\}_{i=1}^{M}$.
    \FOR{each source domain $i$}
        \STATE Compute source reliability $\mathcal{R}_i$ on $\mathcal{B}_s^i$ (Eq.~\eqref{eq:R}).
    \ENDFOR
    \STATE Select hard source: $k_{\text{hard}} \gets \arg\min_i \mathcal{R}_i$
    \STATE Select easy source: $k_{\text{easy}} \gets \arg\max_i \mathcal{R}_i$
    \STATE Update prototypes in $\mathcal{U}_s^{k_{\text{easy}}}$ and $\mathcal{U}_t$ via Eqs.~\eqref{eq:ema_source} and \eqref{eq:ema_target}.
    \STATE \textbf{Step 1:} Update parameters of $f(\cdot)$, $h(\cdot)$, and $d(\cdot)$ using the hard batch $\mathcal{B}_s^{k_{\text{hard}}}$ and $\mathcal{B}_t$ via Eq.~\eqref{eq:main_loss}.
    \STATE \textbf{Step 2:} Update parameters of $g(\cdot)$ using the easy batch $\mathcal{B}_s^{k_{\text{easy}}}$ and $\mathcal{B}_t$ via Eq.~\eqref{eq:clu_loss_easy}.
\ENDFOR
\RETURN Trained HEDN model
\end{algorithmic}
\end{algorithm}

To fully leverage the HEDN architecture while avoiding gradient conflicts between the adversarial alignment and structural refinement objectives, we adopt a step-wise joint optimization strategy. This approach alternates between updating the Hard Network for domain invariance and the Easy Network for structural consistency. The detailed training procedure is summarized in Algorithm~\ref{alg:training}.

\textbf{Step 1: Main Optimization}
In the first step, we optimize the primary objective of the Hard Network using the SRA-selected Hard Source mini-batch and the target mini-batch:
\begin{equation}
\label{eq:main_loss}
\mathcal{L}_{\text{main}} = \mathcal{L}_{\text{cls}}^s + \lambda_1 \mathcal{L}_{\text{adv}} + \lambda_2 \mathcal{L}_{\text{cl}},
\end{equation}
where $\mathcal{L}_{\text{cls}}^s$ enforces emotion-discriminative features on the hard source, $\mathcal{L}_{\text{adv}}$ aligns marginal distributions via DANN, and $\mathcal{L}_{\text{cl}}$ aligns the Hard Network’s predictions with the Easy Network’s pseudo-labels. During this step, we update the shared feature extractor $f(\cdot)$, emotion classifier $h(\cdot)$, and domain discriminator $d(\cdot)$, while keeping $g(\cdot)$ frozen.

\textbf{Step 2: Structural Enhancement via Easy Network.}  
In the second step, we freeze all components from Step 1 and exclusively update the prototype-specific feature extractor $g(\cdot)$ using the cluster-level contrastive loss computed on the SRA-selected Easy Source mini-batch and the target mini-batch:
\begin{equation}
\mathcal{L}_{\text{struct}} = \mathcal{L}_{\text{clu}}^s + \mathcal{L}_{\text{clu}}^t.
\end{equation}
This step refines the semantic embedding space to better conform to the precomputed cluster structure, enhancing intra-cluster compactness and inter-cluster separability without interference from adversarial gradients. By decoupling these objectives temporally, our step-wise strategy preserves the integrity of both domain alignment and structural fidelity.

\subsection{Prediction}

Building on the cluster-based prediction approach of~\cite{wood2024cluster}, we extend this methodology to enable effective cross-subject inference via a structure-aware label transfer strategy. During inference, target samples are classified by leveraging the domain-specific prototype banks maintained for both source and target domains throughout training.

Given a target sample $x^t$, we first extract its semantic feature $\mathbf{z}^t = g(f(x^t))$ and assign it to the nearest target cluster prototype $\mu_{t,k^*} \in \mathcal{U}_t$ based on cosine similarity:
\begin{equation}
    k^* = \arg\max_k \cos(\mathbf{z}^t, \mu_{t,k}).
\end{equation}
Next, for each source domain $i \in \{1, \ldots, M\}$, we identify the source cluster prototype most similar to $\mu_{t,k^*}$:
\begin{equation}
    j^*_i = \arg\max_j \cos(\mu_{t,k^*}, \mu_{i,j}),
\end{equation}
and retrieve its associated emotion label $y_{i, j^*}$. The final predicted label for $x^t$ is determined via majority voting over the labels collected from all source domains:
\begin{equation}
    \hat{y}^t = \text{Vote}(\{y_{i,j^*}\}_{i=1}^M).
\end{equation}
This consensus-based mechanism effectively mitigates label bias or noise from any individual source domain, promoting robust and reliable cross-subject generalization by aggregating structural evidence across multiple reliable prototypes.

\section{Experiment}

\subsection{Datasets}
We evaluate the performance of HEDN on three public EEG emotion datasets: SEED~\cite{zheng2015investigating}, SEED-IV~\cite{zheng2018emotionmeter}, and DEAP~\cite{koelstra2011deap}, all of which record EEG responses to emotion-eliciting video stimuli.

The SEED dataset contains EEG recordings from 15 subjects who viewed 15 Chinese movie clips selected to evoke three emotional states: positive, neutral, and negative. Each subject completed three experimental sessions separated by at least one week. EEG signals were recorded using a 62-channel ESI NeuroScan system. Following prior work, we employ the publicly released Differential Entropy (DE) features provided with the dataset. Each trial was divided into non-overlapping 1-second windows, and DE features were computed across five frequency bands ($\delta$, $\theta$, $\alpha$, $\beta$, and $\gamma$) for all channels, yielding a 310-dimensional feature vector per sample. Subsequently, the features were further processed using a Linear Dynamical System (LDS) projection for temporal smoothing. The final processed dataset contains 3,394 samples per subject in each session.

The SEED-IV dataset also consists of EEG data from 15 subjects, who viewed 24 Chinese movie clips designed to evoke four emotions: happy, sad, fear, and neutral. Similar to SEED, precomputed DE features are used as the model input. A key methodological difference is the segmentation procedure: SEED-IV divides each trial into overlapping 4-second windows. Due to variations in experimental design and stimulus composition across sessions, the effective data length differs slightly among sessions. As a result, each subject contributes 851, 832, and 822 samples in the three respective sessions.

The DEAP dataset includes EEG recordings from 32 subjects who watched 40 one-minute music video clips intended to induce varying emotional responses. Unlike SEED and SEED-IV, DEAP adopts a continuous two-dimensional emotion model in which subjects rate each trial from 1 to 9 on Arousal and Valence. EEG signals were recorded using a 32-channel Biosemi ActiveTwo system. To align DEAP with the discrete classification tasks of SEED and SEED-IV, we binarized the continuous ratings by labeling scores of 5 or higher as high (e.g., high arousal/valence) and those below 5 as low. For the DEAP dataset, EEG trials were segmented into non-overlapping 1-second windows, and DE features were extracted for each window, resulting in 160-dimensional feature vectors. Each subject in DEAP contributes 2,400 samples.

\subsection{Implementation Details}

\begin{table}[!t]
\centering
\caption{Detailed architecture of the modules in HEDN.}
\label{tab:architecture}
\begin{tabularx}{0.97\linewidth}{l l}
\toprule
\textbf{Component} & \textbf{Architecture Flow} \\
\midrule
Feature extractor $f(\cdot)$ & $\mathbf{D_{in}} \to 64 (\text{ReLU}) \to 64 (\text{ReLU})$ \\
Prototype extractor $g(\cdot)$ & $64 \to 32 (\text{ReLU, BN}) \to 64$ \\
Emotion classifier $h(\cdot)$ & $64 \to 32 \to 64 \to \mathbf{C}$ \\
Domain discriminator $d(\cdot)$ & $64 \to 64 (\text{ReLU, Dropout}) \to 1 (\text{Sigmoid})$ \\
\bottomrule
\end{tabularx}
\begin{tablenotes}
\footnotesize
\item All numeric values represent fully connected layer output dimensions. $\mathbf{D_{in}}$ denotes the input dimension (e.g., 310), and $\mathbf{C}$ denotes the number of classes (e.g., 3). BN stands for BatchNorm1d.
\end{tablenotes}
\end{table}

In our experiments, the shared feature extractor $f(\cdot)$, the emotion classifier $h(\cdot)$, the domain discriminator $d(\cdot)$, and the prototype extractor $g(\cdot)$ are all implemented using Multi-Layer Perceptron (MLP) models with the Relu activation function. The layer sizes of each module are summarized in Table~\ref{tab:architecture}.

Regarding hyperparameters, we adapt the RMSprop optimizer with a learning rate of 1e-3 and a weight decay coefficient of 1e-5. The maximum number of training iterations is set to 1000. Due to the differing data scales and experimental settings across datasets, we adopt batch sizes of 96, 96, and 64 for DEAP, SEED, and SEED-IV, respectively. In the main objective loss, the adversarial balancing coefficient $\lambda_1$ follows an exponential growth schedule consistent with the literature~\cite{zhou2023pr}, while the cross-network consistency coefficient $\lambda_2$ is fixed at 0.01. For prototype updates, the source momentum $\gamma_{s}$ is set to 0.5, and the target momentum $\gamma_{t}$ is set to 0.1. To ensure reproducibility, the random seed is fixed at 42. All experiments are implemented using PyTorch and conducted on an NVIDIA RTX 5070Ti GPU.

To determine the DBSCAN clustering configuration, we perform a grid search on the target subject, evaluating combinations of $\epsilon \in [1.0, 5.0]$ and $\text{min\_samples} \in [3, 6]$ using Normalized Mutual Information (NMI) between cluster assignments and trial identifiers. This strategy is motivated by the fact that intra-class structure in EEG datasets often aligns with trial labels. The configuration yielding the highest NMI are retained and applied uniformly across all subjects.

\subsection{Experiment Protocols}

To rigorously evaluate the performance of the proposed HEDN model, we employ two different cross-validation evaluation protocols. 

\textit{(1) Cross-Subject Evaluation.} The cross-subject evaluation protocol is the most widely adopted and fundamental approach for assessing the generalization capability of EEG-based emotion recognition models. In this study, we adopt the widely recognized Leave-One-Subject-Out (LOSO) cross-validation strategy, where data from one subject are held out as the test set, and data from all remaining subjects are used to train the model. This procedure is repeated for all subjects, ensuring that each subject is used as the test set exactly once. 

\textit{(2) Cross-Dataset Evaluation.} While most studies focus on cross-subject evaluation within a single dataset, real-world applications often involve substantial domain shifts due to differences in experimental protocols and environmental conditions. Cross-dataset evaluation assesses the robustness and generalizability of the model across heterogeneous data sources, providing a more realistic measure of performance for practical deployment.

\subsection{Cross-Subject Evaluation}

For the SEED and SEED-IV datasets, experiments are conducted separately on each of the three sessions, resulting in a total of 45 evaluation runs (3 sessions $\times$ 15 subjects). For the DEAP dataset, the model is evaluated on both the Valence and Arousal dimensions, performing 32 runs for each dimension. Reported results include the mean accuracy and standard deviation.

\subsubsection{Evaluation Results on SEED series datasets}

\begin{table}[!t]
\centering
\caption{Performance of HEDN Compared to State-Of-The-Art Approaches on SEED and SEED-IV Datasets.}
\label{tab:seed_results}
\begin{tabularx}{0.9\linewidth}{l C C}
\toprule
    Methods & SEED (\%) & SEED-IV (\%) \\
    \midrule
    DANN~\cite{li2018cross} & 84.93 $\pm$ 08.63 & 76.70 $\pm$ 08.63 \\
    UDDA~\cite{li2022dynamic} & 88.10 $\pm$ 06.54 & 73.14 $\pm$ 09.43 \\
    MS-MDA~\cite{chen2021ms} & 89.63 $\pm$ 06.79 & 59.34 $\pm$ 05.48 \\
    MSDA-SFE~\cite{guo2023multi} & 91.65 $\pm$ 02.91 & 73.92 $\pm$ 06.04 \\
    S$^2$A$^2$-MSDA~\cite{yang2024spectral} & 90.11 $\pm$ 07.32 & 76.23 $\pm$ 09.02 \\
    Gusa~\cite{li2024gusa} & (91.77 $\pm$ 05.91) & 75.12 $\pm$ 07.99 \\
    ADANN~\cite{hong2024adaptive} & 91.92 $\pm$ 05.44 & 73.95 $\pm$ 09.08 \\
    PR-PL~\cite{zhou2023pr} & (93.06 $\pm$ 05.12) & (81.32 $\pm$ 08.53) \\
    TT-CDAN~\cite{huang2025cda} & (93.62 $\pm$ 05.80) & (82.16 $\pm$ 09.67) \\
    GDDN~\cite{chen2024gddn} & 92.54 $\pm$ 03.65 & 75.65 $\pm$ 05.47 \\
    DBPM~\cite{wang2025fine} & \textbf{94.86 $\pm$ 06.43} & \underline{78.90 $\pm$ 09.71} \\
    \midrule
    \multirow{2}{*}{\underline{HEDN (Ours)}} & \underline{94.00 $\pm$ 07.05} & \textbf{79.36 $\pm$ 09.62} \\
    & (96.39 $\pm$ 05.26) & (81.14 $\pm$ 07.77)\\
    \bottomrule
\end{tabularx}
\begin{tablenotes}
\footnotesize
\item The results shown with "()" correspond to the best session results. Bold and underline indicate the best and second-best average performances, respectively.
\end{tablenotes}
\end{table}

To validate the effectiveness of our proposed method on SEED series datasets, we compare HEDN against a comprehensive set of representative EEG emotion recognition approaches. DANN~\cite{li2018cross} and UDDA~\cite{li2022dynamic} serve as a classic single-source domain adaptation baseline, while MS-MDA~\cite{chen2021ms}, MSDA-SFE~\cite{guo2023multi}, and S$^2$A$^2$-MSDA\cite{yang2024spectral} represent standard multi-source domain adaptation methods that construct separate networks for each source-target domain pair. Gusa~\cite{li2024gusa}, ADANN~\cite{hong2024adaptive} and PR-PL~\cite{zhou2023pr} are typical approaches that integrate prototype learning into emotion recognition, and TT-CDAN~\cite{huang2025cda} further enhances domain adaptation through a Temporal Transformer with conditional adversarial strategies. More recently, GDDN~\cite{chen2024gddn} has combined prototype learning with multi-source domain adaptation and further removes source domains with poor transferability to enhance robustness, and DBPM~\cite{wang2025fine}, which achieves reliable cross-subject recognition through fine-grained prototype structures and sequential multi-source training.

As shown in Table~\ref{tab:seed_results}, our proposed HEDN achieves highly competitive performance on both SEED and SEED-IV. On the SEED dataset, HEDN achieves a competitive accuracy of 94.00\%, ranking as the second best among all approaches, closely following DBPM (94.86\%) and significantly outperforming established baselines like ADANN (91.92\%) and GDDN (92.54\%). On the more challenging SEED-IV dataset, HEDN achieves the highest accuracy of 79.36\%, exceeding the second-best baseline DBPM by 0.46\%. These results verify the robustness and strong cross-subject generalization capability of HEDN.

\begin{figure}[tbp]
  \centering
  \includegraphics[width=\linewidth]{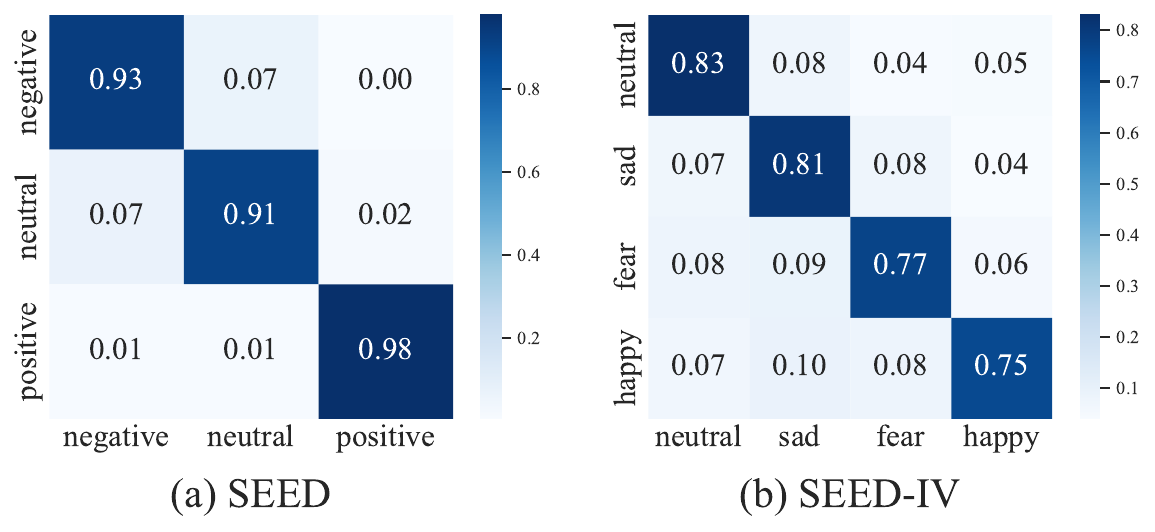}
  \caption{Confusion matrices of HEDN on SEED and SEED-IV datasets.}
  \label{fig:confusion_matrix_for_seed}
\end{figure}

To comprehensively evaluate HEDN's classification performance across different emotional states, we present the confusion matrices for the SEED and SEED-IV datasets in Fig.~\ref{fig:confusion_matrix_for_seed}. On the SEED dataset, HEDN exhibits robust and balanced performance across all three emotion categories, with inter-class confusion consistently below 7\%, demonstrating high accuracy without compromising class balance. On the more challenging four-class SEED-IV dataset, HEDN maintains strong discriminative capability, as most paired misclassification rates remain at or below 10\%, highlighting its effectiveness in distinguishing nuanced emotional states.

\subsubsection{Evaluation Results on DEAP dataset}
Table~\ref{tab:deap_results} summarizes the performance of HEDN on the DEAP dataset. We compare HEDN with several representative baselines, including DANN, MSDA-SFE, and PR-PL, which are widely used in EEG emotion recognition. ATDD-LSTM and DISD-Net are also included, as they are specifically optimized for the DEAP dataset. In addition, DBPM is incorporated as the current state-of-the-art method. Overall, HEDN achieves superior performance across both the valence and arousal dimensions, reaching 73.07\% and 73.20\%, respectively. Notably, HEDN maintains consistently strong performance even when compared with models that have been specially optimized for DEAP (ATDD-LSTM, DISD-Net), demonstrating its generalizability and confirming its robustness and effectiveness in addressing cross-subject variability.

\begin{table}[!t]
\centering
\caption{Performance of HEDN Compared to State-Of-The-Art Approaches on DEAP Dataset.}
\label{tab:deap_results}
\begin{threeparttable}
\begin{tabularx}{0.9\linewidth}{l C C}
\toprule
Methods & Valence (\%) & Arousal (\%) \\
\midrule 
DANN\textsuperscript{*}~\cite{li2018cross} & 69.20 $\pm$ 5.99 & 68.68 $\pm$ 5.63 \\
ATDD-LSTM~\cite{du2020efficient} & 59.06 $\pm$ 6.47 & \underline{72.97 $\pm$ 6.57} \\
MSDA-SFE~\cite{guo2023multi} & 69.26\textsuperscript{$\dagger$} & 70.10\textsuperscript{$\dagger$} \\
ASJDA~\cite{liu2024enhancing} & 68.30 $\pm$ 8.93 & 69.31 $\pm$ 11.26 \\
PR-PL\textsuperscript{*}~\cite{zhou2023pr} & 68.52 $\pm$ 6.43 & 69.22 $\pm$ 7.82 \\
DISD-Net~\cite{cheng2025disd} & 68.24 $\pm$ 6.84 & 71.02 $\pm$ 8.70 \\
DBPM\textsuperscript{*}~\cite{wang2025fine} & \underline{70.65 $\pm$ 6.62} & 71.20 $\pm$ 6.24 \\
\midrule
HEDN & \textbf{73.07 $\pm$ 05.01} & \textbf{ 73.20 $\pm$ 08.97} \\
\bottomrule
\end{tabularx}
\begin{tablenotes}
    \footnotesize
    \item[$\dagger$] The standard deviation not reported in the original publication.
    \item[*] The corresponding results are reproduced in this work.
\end{tablenotes}
\end{threeparttable}
\end{table}

\begin{figure}[!t]
  \centering
  \includegraphics[width=\linewidth]{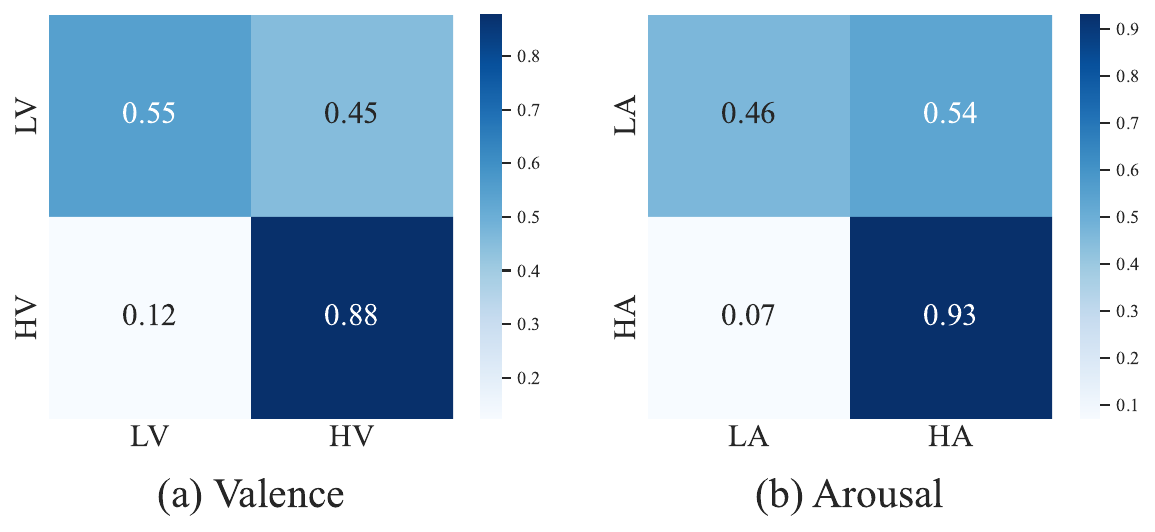}
  \caption{Confusion matrices of HEDN on DEAP dataset.}
  \label{fig:confusion_matrix_for_deap}
\end{figure}

Figure~\ref{fig:confusion_matrix_for_deap} shows the confusion matrices for the DEAP dataset. For both valence and arousal dimensions, high states are generally easier to classify, whereas low states are more prone to misclassification, with nearly half of low samples predicted as high. This pattern reflects the inherent class imbalance in DEAP. Future work focuses on developing specialized algorithms to address this imbalance and improve performance on underrepresented states.

\subsection{Cross-Dataset Evaluation}
To systematically assess the cross-dataset generalization of our proposed method, we conduct two transfer experiments: $\text{SEED} \to \text{SEED-IV}$ and $\text{SEED-IV} \to \text{SEED}$. Due to differences in emotion label distributions between the two datasets, we first perform class alignment. Specifically, the “sad” and “fear” classes in $\text{SEED-IV}$ are merged into a single class corresponding to the “negative” category in $\text{SEED}$, while the “neutral” and “happy” classes are mapped to the “neutral” and “positive” categories in $\text{SEED}$, respectively. This procedure resulted in a three-class classification task, allowing for a consistent evaluation of our method across both datasets.
 
For the experimental protocol, we adopt a widely accepted Subject-Level cross-dataset evaluation setup~\cite{lan2018domain, zhou2025enhancing}: all subjects in the source dataset are used for training, and one subject from the target dataset is used for testing. This process is repeated until all subjects in the target dataset have been evaluated once, and the average accuracy is reported.

Table~\ref{tab:cross_dataset_results} summarizes the cross-dataset performance of HEDN. HEDN achieves 69.69\% accuracy on $\text{SEED} \to \text{SEED-IV}$ and 73.63\% on $\text{SEED-IV} \to \text{SEED}$, outperforming all existing methods and establishing a new state of the art. These results demonstrate that HEDN’s SRA-driven dynamic routing and dual-branch architecture effectively extract transferable knowledge from heterogeneous sources, enabling robust cross-dataset emotion recognition.

\begin{table}[!t]
\centering
\caption{Performance of HEDN Compared to State-Of-The-Art Approaches on Cross-Dataset Tasks.}
\label{tab:cross_dataset_results}
\begin{tabularx}{\linewidth}{l C C}
\toprule
Methods & $\text{SEED} \to \text{SEED-IV}$ (\%) & $\text{SEED-IV} \to \text{SEED}$ (\%) \\
\midrule 
IAG~\cite{song2020instance}  & 49.97 & 51.98 \\
GMSS~\cite{li2022gmss}   & 55.72 & 53.10 \\
DAN~\cite{li2018cross} & 55.56 & 48.18 \\
MSDA-SFE~\cite{guo2023multi} & 62.10 & 59.15 \\
GDDN~\cite{chen2024gddn}  & 63.07 & 59.15 \\
E$^2$STN~\cite{zhou2025enhancing} & 61.62 & 60.51 \\
DBPM~\cite{wang2025fine} & 65.95 & 70.59 \\
\midrule
HEDN & \textbf{69.69} & \textbf{73.63} \\
\bottomrule
\end{tabularx}
\end{table}

\section{DISCUSSION}

\subsection{Ablation Study}
\begin{table}[!t]
  \centering
  \caption{Ablation study of HEDN on SEED and SEED-IV datasets.}
  \begin{tabularx}{\linewidth}{lcc}
    \toprule
    Methods & SEED (\%) & SEED-IV (\%) \\
    \midrule
    \multicolumn{3}{l}{\textbf{Ablation study about SRA routing}} \\
    \midrule
    w/ Random Routing & 92.20$\pm$06.93 & 78.71$\pm$09.10 \\
    w/ Easy Source in Both Networks & 81.50$\pm$11.21 & 57.78$\pm$08.00 \\
    w/ Hard Source in Both Networks & 92.83$\pm$06.70 & 79.10$\pm$09.77 \\
    \midrule
    \multicolumn{3}{l}{\textbf{Ablation study about HEDN components}} \\
    \midrule
    w/o Hard Network & 79.25$\pm$11.92 & 63.77$\pm$09.87 \\
    w/o Easy Network & 85.19$\pm$08.65 & 77.11$\pm$10.31 \\ 
    w/o $\mathcal{L}_{clu}$ on the Source & 93.86$\pm$07.04 & 78.04$\pm$09.88 \\
    w/o $\mathcal{L}_{clu}$ on the Target & 90.71$\pm$07.09 & 79.09$\pm$08.89 \\
    w/o $\mathcal{L}_{clu}$ on the Source and Target & 89.45$\pm$07.61 & 78.81$\pm$09.78 \\
    w/o Step-Wise Joint Optimization & 90.22$\pm$10.02 & 74.32$\pm$09.56 \\
    \midrule
    \textbf{HEDN} & \textbf{94.00$\pm$07.05}  & \textbf{79.36$\pm$09.62} \\
    \bottomrule 
  \end{tabularx}
\label{tab:ablation_study}
\end{table}

To further investigate the contributions of individual components in our proposed method, we conduct a series of ablation experiments focusing on two key aspects: the SRA routing mechanism, and architectural and objective-level components within the HEDN. The results are summarized in Table~\ref{tab:ablation_study}.

We first evaluate the importance of dynamic source routing by comparing three alternatives: (1) w/ Random Routing, where source domains are randomly assigned to the Easy or Hard networks regardless of their difficulty scores. Its inferior performance compared to HEDN shows that the effectiveness of our framework comes from SRA’s ability to assess intrinsic transferability, not from the dual-branch architecture alone; (2) w/ Easy Source in Both Networks, which indicates that sources categorized as Easy Source are simultaneously fed into both the Easy and Hard networks. This strategy causes Easy sources to be over-processed, which leads the model to overlook more challenging source domains and ultimately degrades performance; and (3) w/ Hard Source in Both Networks, which indicates that sources categorized as Hard Source are simultaneously fed into both networks. This strategy causes the Easy Network cannot effectively propagate labels from high-quality sources to the target domain. Consequently, the target domain fails to obtain reliable pseudo-labels, resulting in reduced accuracy.

For the architectural and objective-level components of HEDN, we examine six aspects. (1) w/o Hard Network removes the Hard Network entirely, leading to a substantial accuracy drop of more than 10\% on both datasets, since the Hard Network is essential for mitigating the distribution discrepancy between source and target domains. (2) w/o Easy Network removes the Easy Network, resulting in the accuracy on the SEED dataset decreasing to 85.19\%, which highlights its critical role in label propagation. (3) w/o $\mathcal{L}_{clu}$ on the Source, (4) w/o $\mathcal{L}_{clu}$ on the Target, and (5) w/o $\mathcal{L}_{clu}$ on the Source and Target correspond to the ablation of Eq. (13). The results show that removing $\mathcal{L}_{clu}$ entirely yields the largest accuracy degradation, and $\mathcal{L}_{clu}$ on the target domain exhibits a greater influence on the SEED dataset. Moreover, $\mathcal{L}_{clu}$ is evidently more crucial on SEED than on SEED-IV; on SEED-IV, removing it only reduces the accuracy from 79.36\% to 78.81\%, yielding a small degradation of 0.55\%, which suggests that the cluster structure in SEED-IV is more stable and less sensitive to training perturbations. (6) w/o Step-Wise Joint Optimization, which optimizes all objectives in a single step, causes even greater performance degradation than removing the second stage (i.e., w/o $\mathcal{L}_{clu}$ on both the Source and Target) on the SEED-IV dataset, indicating that the two-step optimization scheme, which directs the model to focus on distinct tasks at different stages, helps maintain stable cluster structures and facilitates reliable label propagation.

Overall, these results confirm that each component of HEDN is essential. SRA facilitates intelligent source selection, the dual-branch architecture accommodates heterogeneous transfer requirements, and the step-wise contrastive learning preserves structural integrity, collectively supporting robust cross-subject emotion recognition.

\subsection{Complexity Analysis}

\begin{table*}[!t]
\centering
\caption{Comparison of Computational Efficiency Across Different Methods}
\label{tab:comp_ana}
\begin{threeparttable}
\begin{tabularx}{0.9\linewidth}{lcccccc}
    \toprule
    \textbf{Method} & \textbf{FLOPs (M)} & \textbf{Params} & \textbf{Model Size (KB)} & \textbf{Adaptation Time\textsuperscript{*} (s)} & \textbf{Inference Time\textsuperscript{*} (ms)} & \textbf{Accuracy\textsuperscript{\dag} (\%) } \\
    \midrule
    DANN & \textbf{5.81} & 32644 & \textbf{137.57} & 140.58 & 1.93 & 84.93 \\
    MSMDA & 34.99 & 150883 & 649.42 & 843.94 & 4.81 & 85.26 \\ 
    S$^2$A$^2$-MSDA & 65.25 & 156458 & 670.83 & 7768.20 & 6.26 & 87.84\textsuperscript{\ddag} \\
    PR-PL & 7.69 & \textbf{28289} & 137.58 & 432.90 & 2.78 & 89.89\\
    DBPM & 6.22 & 32676 & 138.93 & 190.91 & \textbf{1.91} & 93.83 \\ 
    HEDN & 61.55 & 36932 & 219.14 & \textbf{48.98} & 9.72 & \textbf{94.00} \\
    \bottomrule
\end{tabularx}
\begin{tablenotes}
    \footnotesize
    \item[*] Adaptation and inference times are measured on the first session of the SEED dataset under identical hardware conditions with a single process, averaged over all target subjects.
    \item[$\dag$] Accuracy is reproduced using random seed 42, with all other settings consistent with the original publications.
    \item[$\ddag$] Due to the prohibitive training cost of S$^2$A$^2$-MSDA, an early stopping strategy with patience of 1,000 iterations is applied.
\end{tablenotes}
\end{threeparttable}
\end{table*}

Wearable brain–computer interfaces require models that are both accurate and computationally efficient, with the ability to be rapidly deployed. This requirement is especially critical for cross-subject emotion recognition with domain adaptation, where adaptation relies on unlabeled target EEG data. To support real-time operation, such models must complete adaptation promptly upon receiving user data, making fast adaptation essential.

To evaluate the efficiency of the proposed HEDN model, we compare it with several publicly available baselines from Table~\ref{tab:seed_results} that provide open-source implementations. As shown in Table~\ref{tab:comp_ana}, HEDN achieves high accuracy while substantially improving adaptation efficiency. Specifically, HEDN requires only 48.98 seconds for adaptation, which is more than four times faster than the next-best competitor and up to 160 times faster than the slowest baseline. This efficiency derives from its SRA dynamic routing: at each iteration, HEDN selects only the most and least reliable source domains and routes them to two specialized branches. By selectively processing informative samples rather than uniformly traversing all source data, HEDN eliminates redundant computations, maximizes information gain per update, and reduces overall computational burden.

Although the per-iteration SRA evaluation increases the FLOPs count (61.55M) compared to other methods, HEDN converges rapidly and remains compact, with only 36,932 parameters and a model size of 219.14 KB. Its inference latency of 9.72 ms is well within acceptable limits for real-time EEG processing, particularly considering the substantial improvements in both adaptation speed and accuracy. Consequently, HEDN achieves an optimal balance between efficiency and performance: it avoids the high adaptation costs of complex multi-source methods (e.g., MSMDA, S$^2$A$^2$-MSDA) while outperforming single-source approaches (e.g., DANN, PR-PL) in both speed and accuracy.

In summary, HEDN not only excels in emotion recognition performance but also meets the stringent real-time and resource constraints of wearable BCI applications, making it a highly promising solution for practical deployment.

\subsection{The effect of SRA mechanism}
\label{sec:tda_effect}
\begin{figure}[!t]
    \centering
    \includegraphics[width=\linewidth]{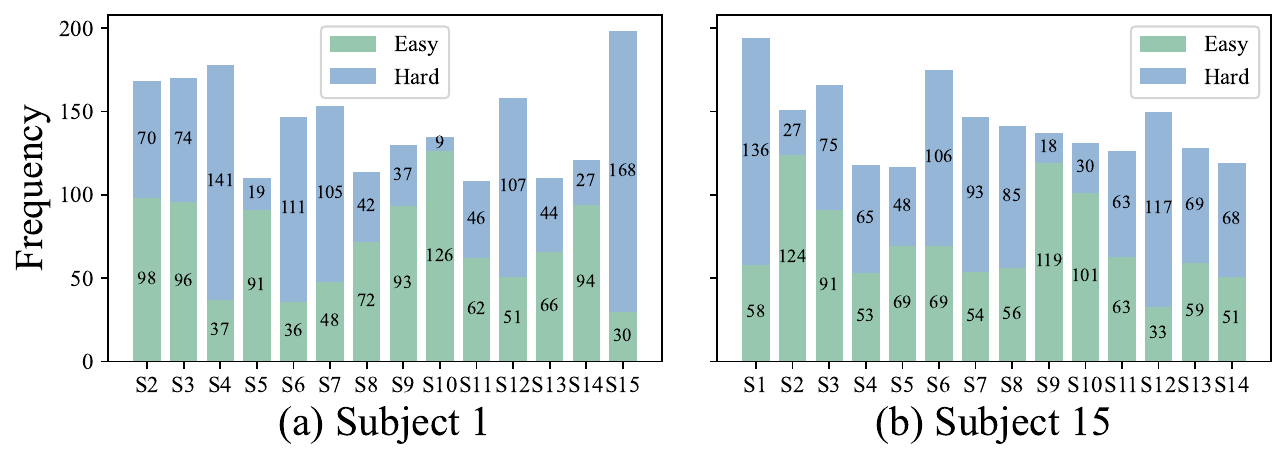}
    \setlength{\abovecaptionskip}{-0.3cm}
    \caption{Selection frequency of each source domain over the entire training process.}
    \label{fig:frequency}
\end{figure}

To further evaluate the effectiveness of the proposed Source Reliability Assessment (SRA) mechanism, we analyze the frequency with which each source domain is selected as a Hard or Easy Source during training. As shown in Fig.~\ref{fig:frequency}, sources such as S9 and S10 are consistently selected as Easy Source across different target subjects, indicating that they possess stable internal structures independent of the target domain. Nevertheless, the selection frequency of certain sources varies with the target, for example, S2 is selected as an Easy Source far more often when the target is Subject 15 than when it is Subject 1. Although SRA itself does not use any target domain information, this variation reflects an interaction between intrinsic source reliability and cross-domain alignment. Specifically, even a structurally reliable source may be treated as low-quality in practice if its marginal distribution differs significantly from that of the target; during adversarial alignment, the Hard Network’s attempts to match distributions can distort the source’s internal cluster structure. Consequently, while SRA does not directly measure source–target similarity, the overall routing behavior implicitly captures distributional compatibility through the adaptation process.

\begin{figure}[!t]
    \centering
    \includegraphics[width=\linewidth]{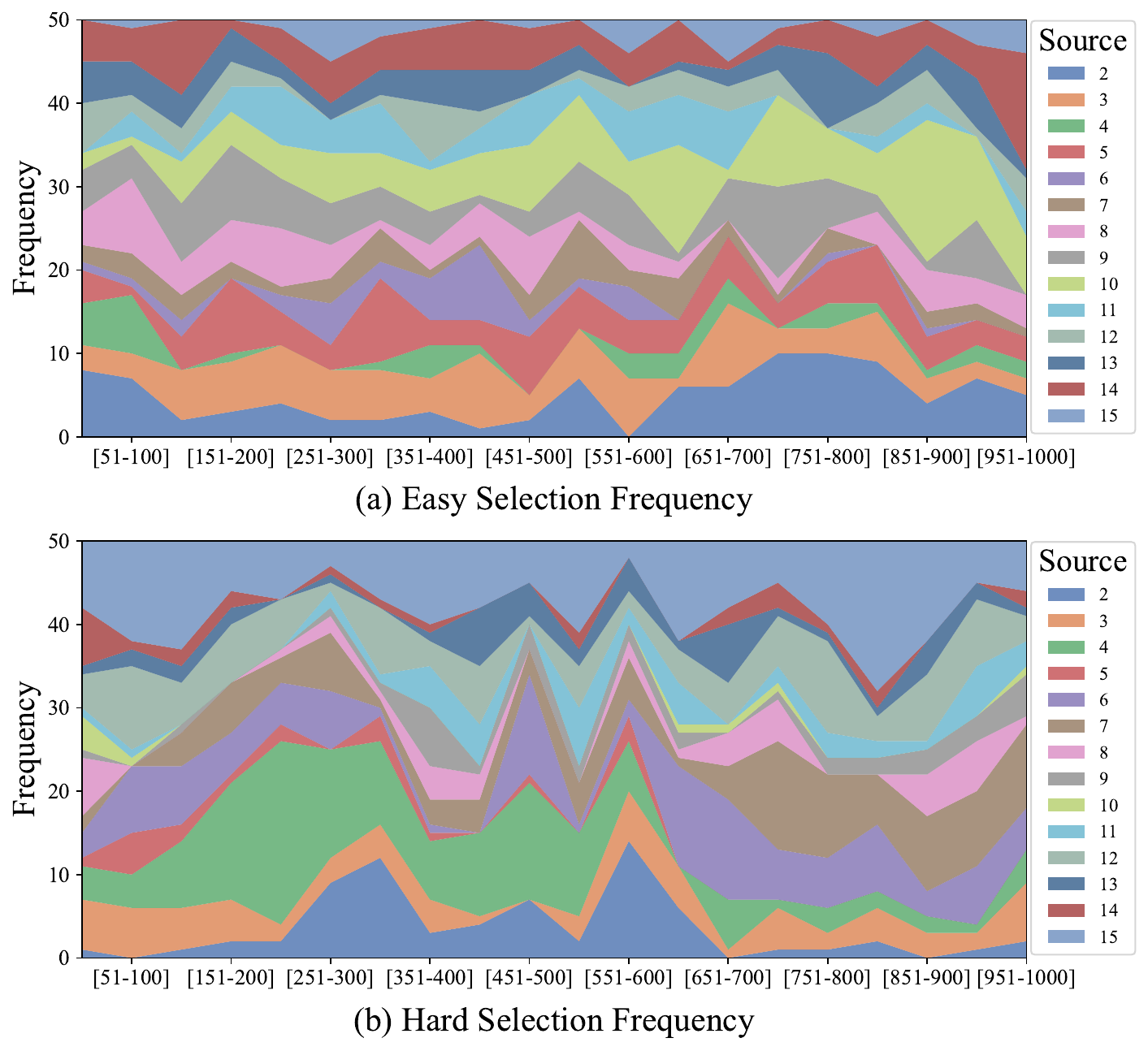}
    \setlength{\abovecaptionskip}{-0.3cm}
    \caption{Frequency distribution of source selection. (a) Easy sources selection frequency; (b) Hard sources selection frequency.}
    \label{fig:time_frequency}
\end{figure}

While the above analysis offers a global perspective, it does not reflect the dynamic nature of source selection during training. To reveal how the SRA mechanism evolves, Figure~\ref{fig:time_frequency} presents the proportions of sources selected as Easy or Hard Source at different training stages, using the first subject in the SEED dataset as the target domain. The selection of Easy Source remains relatively stable across iterations, indicating consistent structural quality. In contrast, Hard Source selection displays clear stage-wise fluctuations, suggesting that the model adaptively prioritizes different low-reliability sources as feature representations mature. Additionally, certain sources switch roles during training. For example, Source 2 is initially processed by the Hard Network but later transitions to the Easy Network, demonstrating that SRA can reassess source reliability as adaptation progresses. This adaptive routing highlights a key advantage of HEDN. Instead of treating reliability as a fixed attribute, the framework interprets it as a context-dependent signal that supports efficient and robust cross-subject knowledge transfer.

\subsection{The effect of hyperparameter settings}

\begin{figure}[!t]
    \centering
    \includegraphics[width=\linewidth]{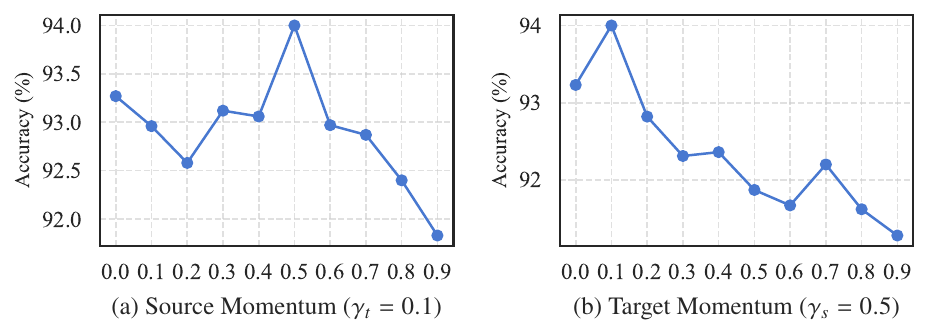}
    \setlength{\abovecaptionskip}{-0.3cm}
    \caption{
        Ablation study on momentum parameters. 
        (a) Source momentum ablation with target momentum fixed at $\gamma_t = 0.1$.
        (b) Target momentum ablation with source momentum fixed at $\gamma_s = 0.5$.}
    \label{fig:params}
\end{figure}

In this section, we analyze the sensitivity of HEDN to the source momentum coefficient $\gamma_s$ and the target momentum coefficient $\gamma_t$, which determine the contribution of historical prototypes in exponential moving average updates. Figure~\ref{fig:params} summarizes the ablation results and shows that the optimal configuration is $\gamma_s = 0.5$ and $\gamma_t = 0.1$. This difference arises from the distinct update dynamics of the two domains.

As discussed in Section~\ref{sec:tda_effect}, the SRA mechanism dynamically selects source domains based on estimated reliability, meaning that each source only participates intermittently during training. Furthermore, each source maintains its own prototype, which is updated only when that source is selected. When a source is skipped for several iterations, its prototype becomes outdated. Therefore, $\gamma_s$ must strike a balance between two competing priorities: retaining enough historical information to prevent catastrophic forgetting during inactive periods while remaining adaptable when the source reappears. The value $\gamma_s = 0.5$ achieves this balance. Larger values overweight outdated historical information and slow adaptation, whereas smaller values make prototypes excessively sensitive to transient updates, leading to instability.

In contrast, the target domain is updated continuously, as target samples appear in every iteration, and its prototype directly influences pseudo-label generation. Accordingly, a lower momentum value, with $\gamma_t = 0.1$ performing best, allows the model to adjust rapidly to evolving target representations while maintaining stability in pseudo-label assignment.

\subsection{Stability Analysis}

The stability of a model under different training conditions is a key factor in measuring its reliability in real-world BCI systems. We evaluate the stability of the HEDN framework along two dimensions: generalization robustness and convergence stability.

\subsubsection{Generalization Robustness}

\begin{table*}[!t]
\centering
\caption{Comparison of Generalization Robustness for HEDN and Baselines on SEED and SEED-IV Datasets.}
\label{tab:gen_rob}
\begin{tabular}{lccccccccc}
\toprule & \textbf{Method} & \textbf{40} (\%) & \textbf{41} (\%) & \textbf{42} (\%) & \textbf{43} (\%) & \textbf{44} (\%) & \textbf{Mean} (\%) & \textbf{Std} (\%) & \textbf{Max-Min} (\%)\\
\midrule
\multirow{6}{*}{SEED}
& DANN      & 84.61 & 84.45 & 84.93 & 84.56 & 84.14 & 84.54 & \textbf{0.28} & \textbf{0.79} \\
& MSMDA     & 86.98 & 86.49 & 85.26 & 85.42 & 85.85 & 86.00 & 0.73 & 1.72 \\
& S$^2$A$^2$-MSDA & 88.25 & 87.60 & 87.84 & 88.16 & 88.96 & 88.16 & 0.51 & 1.36 \\
& PR-PL      & 88.55 & 90.38 & 89.89 & 89.30 & 91.12 & 89.85 & 0.99 & 2.57 \\
& DBPM      & 92.02 & 93.25 & 93.83 & \textbf{94.35} & \textbf{93.54} & \textbf{93.40} & 0.87 & 2.33 \\
& HEDN      & \textbf{92.48} & \textbf{93.95} & \textbf{94.00} & 92.28 & 93.47 & 93.24 & 0.81 & 1.72 \\
\midrule
\multirow{6}{*}{SEED-IV}
& DANN      & 77.80 & 77.47 & 76.70 & 76.69 & 76.63 & 77.06 & 0.54 & 1.17 \\
& MSMDA     & 74.88 & 74.52 & 74.02 & 73.74 & 74.04 & 74.24 & 0.46 & 1.14 \\
& S$^2$A$^2$-MSDA & 78.75 & 77.49 & 77.99 & 76.58 & 78.75 & 77.91 & 0.92 & 2.17 \\
& PR-PL      & 75.08 & 75.31 & 74.47 & 73.93 & 74.69 & 74.70 & 0.54 & 1.38\\
& DBPM      & 77.96 & 78.61 & \textbf{79.38} & 77.14 & 78.35 & 78.29 & 0.82 & 2.24 \\
& HEDN      & \textbf{78.93} & \textbf{79.48} & 79.36 & \textbf{78.86} & \textbf{79.92} & \textbf{79.31} & \textbf{0.43} & \textbf{1.06} \\
\bottomrule
\end{tabular}
\end{table*}

To assess the reliability of HEDN under stochastic variation, we examine model performance across five random seeds (40–44). As shown in Table \ref{tab:gen_rob}, HEDN demonstrates consistently strong performance on both SEED and SEED-IV datasets. On SEED, HEDN achieves a mean accuracy of 93.24\%, closely matching DBPM (93.40\%) while exhibiting a lower variance (Std = 0.81) and a reduced performance fluctuation range (Max–Min = 1.72). This indicates that HEDN delivers highly stable performance despite having a smaller model capacity than DBPM.

The advantage becomes more pronounced on the more challenging SEED-IV dataset. HEDN attains the highest mean accuracy (79.31\%) among all methods while also achieving the lowest standard deviation (0.43) and smallest Max–Min range (1.06).  This highlights the stabilizing effect of HEDN’s SRA-based source selection and dual-branch design, which reduce sensitivity to initialization and stochastic training dynamics.

Overall, HEDN exhibits excellent generalization robustness: it delivers high accuracy with consistently low variance across datasets, confirming that its performance is reliable and not dependent on favorable initialization, making it well-suited for practical BCI deployment.

\subsubsection{Convergence Stability}

\begin{table}[!t]
\centering
\caption{Comparison of Best and Final Accuracies for HEDN and Baselines on SEED and SEED-IV Datasets.}
\label{tab:conv_stab}
\begin{tabularx}{\linewidth}{lccccc}
\toprule
& \textbf{Method} & \textbf{Best (\%)} & \textbf{Final (\%)} & $\Delta_\text{acc}$ (\%) \\
\midrule
\multirow{6}{*}{SEED}
                      & DANN & 84.93 & 76.22 & 8.71 \\
                      & MSMDA & 85.26 & 77.21 & \textbf{8.05} \\
                      & S$^2$A$^2$-MSDA & 87.84 & 79.10 & 8.74 \\
                      & PR-PL & 89.89 & 77.70 & 12.19 \\
                      & DBPM & 93.83 & 77.66 & 16.17 \\
                      & HEDN & \textbf{94.00} & \textbf{84.17} & 9.83 \\
\midrule
\multirow{6}{*}{SEED-IV} & DANN & 76.70 & 56.78 & 19.92 \\
                      & MSMDA & 74.02 & 59.71 & \textbf{14.31} \\
                      & S$^2$A$^2$-MSDA & 77.99 & \textbf{61.43} & 16.56 \\
                      & PR-PL & 74.47 & 55.00 & 19.47 \\
                      & DBPM & \textbf{79.38} & 58.32 & 21.06 \\
                      & HEDN & 79.36 & 59.04 & 20.32 \\
\bottomrule
\end{tabularx}
\end{table}

To assess the convergence stability of HEDN during training, we compare the best accuracy observed throughout training with the final accuracy at convergence. Prior work typically reports only the best recorded accuracy, which may overestimate real-world performance since the optimal checkpoint cannot be reliably identified without target-domain labels.

To provide a more reliable evaluation, we introduce the metric $\Delta_{\text{acc}} =$ Best Accuracy – Final Accuracy, where a smaller value of $\Delta_{\text{acc}}$ indicates that the model maintains strong performance until the final training stage, reflecting greater robustness and stability.

As shown in Table~\ref{tab:conv_stab}, HEDN achieves the highest final accuracy on the SEED dataset (84.17\%) with a moderate gap of $\Delta_{\text{acc}} = 9.83\%$, demonstrating that it not only attains state-of-the-art peak performance but also sustains a high level of accuracy at convergence. On the more challenging SEED-IV dataset, HEDN exhibits a larger gap $\Delta_\text{acc} = 20.32\%$, but its final accuracy (59.04\%) remains competitive, second only to S$^2$A$^2$-MSDA (61.43\%) and MSMDA (59.71\%).

Notably, a clear trend emerges when examining methods by their modeling capacity: multi-source domain adaptation approaches such as MSMDA and S$^2$A$^2$-MSDA exhibit consistently smaller accuracy gaps, indicating smoother convergence behavior due to their explicit multi-source modeling. HEDN, leveraging its dual-branch architecture and SRA mechanism, achieves a favorable balance between performance and stability, outperforming single-source adaptation baselines while avoiding the heavy computational overhead of full multi-source approaches.

\subsection{Source Number Evaluation}
\begin{table}[!t]
\centering
\caption{Effect of source domain number on HEDN performance and efficiency.}
\label{tab:source_number}
\begin{tabularx}{0.9\linewidth}{cccc}
\toprule
\textbf{Sources} &
\makecell{\textbf{Accuracy} (\%)} &
\makecell{\textbf{FLOPs} (M)} &
\makecell{\textbf{Adaptation Time} (s)} \\
\midrule
14 & 94.00 & 61.55 & 48.98 \\
12 & 93.05 & 54.53 & 42.31 \\
10 & 91.91 & 47.52 & 40.41 \\
8  & 91.59 & 40.50 & 32.34 \\ 
6  & 90.56 & 33.48 & 29.10 \\ 
4  & 89.46 & 26.47 & 26.52 \\
2  & 84.75 & 19.45 & 22.76 \\ 
\bottomrule
\end{tabularx}
\end{table}

The number of source domains is a critical factor in multi-source domain adaptation, influencing both predictive performance and computational efficiency. To investigate this trade-off, we evaluate HEDN on the SEED dataset while varying the number of source domains from 2 to 14.

As shown in Table~\ref{tab:source_number}, classification accuracy improves monotonically with the number of sources, rising from 84.75\% with 2 sources to 94.00\% with 14 sources. The most significant gains occur when expanding from 2 to 6 sources, resulting in a gain of 5.81\%. Beyond 8 sources, however, the improvement becomes marginal, suggesting diminishing returns as additional source domains are incorporated.

From an efficiency perspective, both computational cost and adaptation time increase with the number of sources. The FLOPs grow from 19.45M with 2 sources to 61.55M with 14 sources, and the total adaptation time increases from 22.76 seconds to 48.98 seconds. Nevertheless, even in the full 14-source setting, HEDN remains computationally efficient, requiring fewer than 65M FLOPs and less than 50 seconds for adaptation on the SEED dataset. These results demonstrate that HEDN offers a practical and scalable solution for real-world EEG emotion recognition. Users can adjust the number of source domains to balance accuracy and resource constraints, making the framework adaptable to diverse deployment scenarios—from lightweight edge devices to high-performance systems.

\subsection{Impact of Density-Based Prototype Initialization}

\begin{table}[!t]
\centering
\caption{Comparison of prototype initialization strategies on SEED and SEED-IV datasets.}
\label{tab:dbscan}
\begin{tabularx}{\linewidth}{lCC}
\toprule
\textbf{Initialization Method} & \textbf{SEED} (\%) & \textbf{SEED-IV} (\%) \\
\midrule
Trial-based Initialization & 94.00 & 78.50 \\
DBSCAN (Silhouette Tuning) & 92.19 & 78.48 \\ 
DBSCAN (NMI Tuning) & 94.00 & 79.36 \\ 
\bottomrule
\end{tabularx}
\end{table}

To examine the importance of density-based clustering for prototype initialization, we evaluate DBSCAN-based initialization and compare it with using trial identifiers as direct cluster labels. Although emotional EEG datasets often exhibit alignment between latent feature clusters and trial structure, relying solely on trial labels introduces limitations. Trial boundaries do not necessarily reflect the true structure of neural responses, especially during long-duration emotional stimuli where transient emotional drift may occur. Such drift creates scattered or inconsistent samples that behave as outliers. When trial labels are used as cluster assignments, these samples are forced into a single homogeneous category, preventing the prototype from accurately capturing the underlying emotional dynamics.

This effect is evident in the SEED-IV dataset. As shown in Table~\ref{tab:dbscan}, directly using trial labels yields an accuracy of 78.50\%, while DBSCAN with NMI-guided parameter tuning improves performance to 79.36\%. These results demonstrate that density-based clustering with auxiliary structural guidance more effectively preserves latent feature organization. 

Nevertheless, trial labels provide meaningful supervisory information. Rather than treating them as fixed clustering assignments, we employ them as auxiliary guidance to tune DBSCAN parameters via NMI. This strategy leverages the implicit structure present in trials while mitigating sensitivity to noise and outliers.

In scenarios where trial labels are unavailable, DBSCAN remains applicable by tuning parameters using the silhouette score. Although this purely unsupervised criterion yields slightly lower performance (Table~\ref{tab:dbscan}), the accuracy drop is minor and the result remains competitive. It is worth noting that many existing studies assume access to trial information or explicitly utilize trial structure, either through preprocessing pipelines (e.g., trial-wise LDS smoothing) or through trial-aware augmentation strategies, such as those used in EEGMatch~\cite{zhou2024eegmatch}. In this context, tuning DBSCAN using NMI provides a more aligned initialization strategy, balancing structural guidance with robustness to irregular or noisy samples.

Finally, despite requiring hyperparameter search, the computational cost remains negligible. Owing to the modest sample size and low feature dimensionality typical of EEG datasets, DBSCAN tuning completes in approximately 1.8 seconds on average, making the method practical for real-world deployment.

\subsection{Visualization Analysis}

\begin{figure*}[!t]
    \centering
    \includegraphics[width=\linewidth]{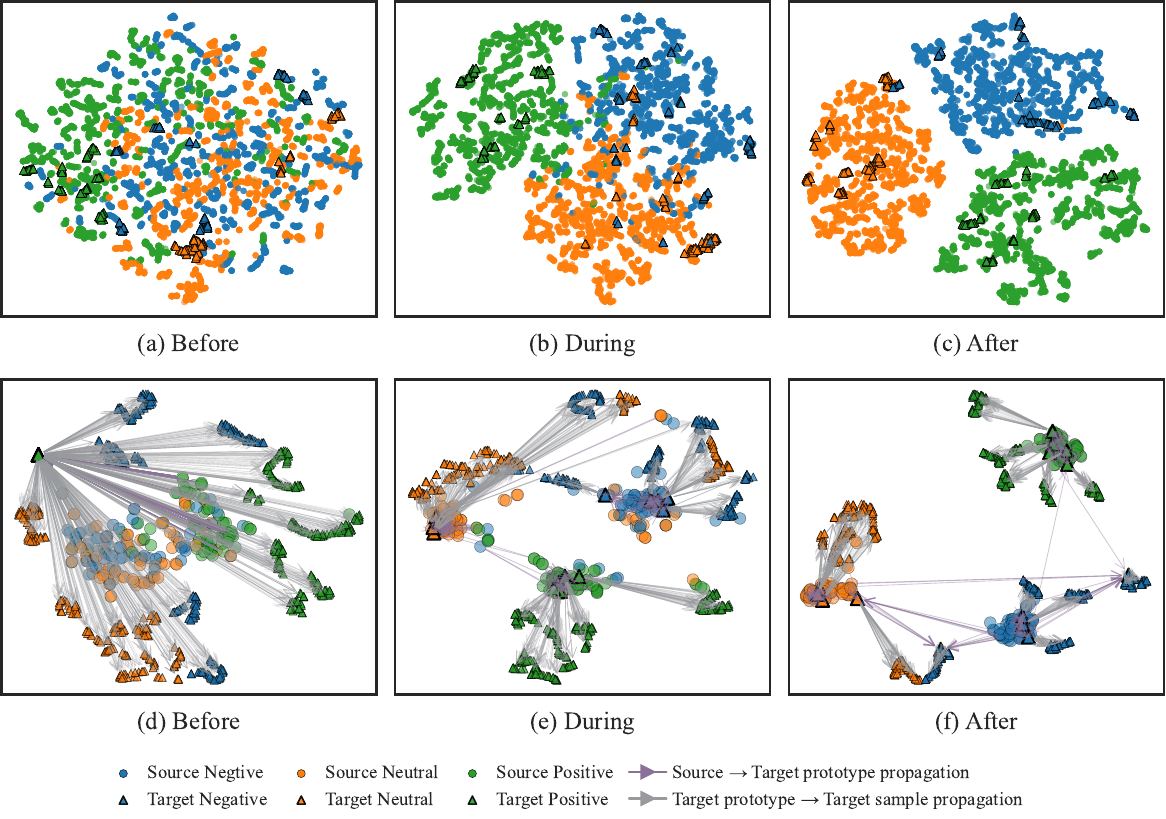}
    \caption{t-SNE visualizations of feature distributions and structure-aware label propagation during cross-subject adaptation. Top row (a–c): t-SNE projections of emotion-related representations at different training stages—before adaptation, during adaptation, and after convergence. Circle markers indicate source samples, and triangle markers indicate target samples. Colors denote emotion categories (negative, neutral, positive). Bottom row (d–f): Visualization of prototype-based label propagation. Source and target cluster prototypes are shown together with a subset of target samples. Arrows illustrate the structure-aware knowledge transfer: source$\rightarrow$target prototype matching (purple arrows) and target-prototype$\rightarrow$target-sample assignment (gray arrows). }
    \label{fig:vis}
\end{figure*}

To qualitatively validate the effectiveness of our HEDN framework, we provide t-SNE visualizations of both the feature distributions and the prototype-based label propagation across three representative training stages: before adaptation, during adaptation, and after convergence. The results are presented in Fig.~\ref{fig:vis}.

In the top row, source subjects and the target subject are color-coded by ground-truth emotion labels (negative, neutral, positive). Before adaptation, features from source and target domains are largely disjoint, reflecting substantial cross-subject variability. During training, the model progressively aligns the feature distributions while preserving intra-class compactness. After convergence, features from the same emotion class cluster tightly together irrespective of domain, demonstrating successful learning of domain-invariant representations.

The bottom row illustrates the mechanics of our inference strategy: structure-aware label propagation via prototype matching. Each target prototype (larger triangle markers) is aligned with multiple source prototypes (larger circle markers) from different subjects through cosine similarity (purple arrows). Prototypes located near decision boundaries often match source prototypes belonging to different emotion categories. Despite this ambiguity, the majority-voting strategy, which aggregates labels from all matched source prototypes, effectively reduces noise and yields reliable pseudo-labels for the target prototypes.

Finally, each target sample is assigned to its nearest target prototype (gray arrows), and the pseudo-label of that prototype serves as the final prediction. This two-level propagation process, comprising (i) cross-domain prototype alignment with robust voting and (ii) sample-to-prototype assignment, ensures that predictions are grounded in both structural consistency and semantic consensus across source domains. Notably, within each emotion class, fine-grained sub-cluster structures become increasingly distinct and stable, suggesting the effectiveness of cluster contrastive learning.

Overall, the visualization confirms that our approach successfully bridges the domain gap at the prototype level while preserving the intrinsic cluster structure of emotional EEG responses, thereby supporting reliable and interpretable cross-subject prediction.

\subsection{Limitations and Future Work}

\begin{figure}[!t]
    \centering
    \includegraphics[width=\linewidth]{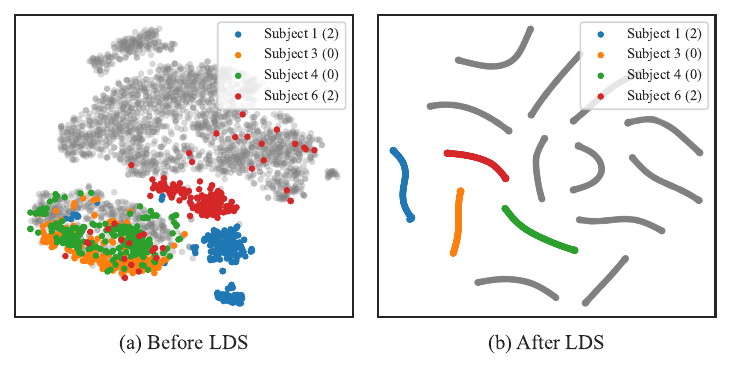}
    \setlength{\abovecaptionskip}{-0.3cm}
    \caption{t-SNE visualization of EEG feature distributions. (a) Raw EEG features before LDS preprocessing, showing scattered and overlapping clusters. (b) Features after LDS preprocessing, where intra-class clusters are more compact and distinguishable, facilitating reliable prototype construction.}
    \label{fig:pre-lds-clusters}
\end{figure}

Although HEDN demonstrates strong performance by leveraging latent intra-class cluster structures for prototype-based prediction, its effectiveness remains constrained by the clarity and separability of these clusters. Stable clusters that enable reliable prototype construction primarily emerge after LDS preprocessing. While multiple subclusters are discernible in the raw EEG space, they are more diffuse and difficult to model robustly, as illustrated in Fig.~\ref{fig:pre-lds-clusters}. Without LDS, the scattered and overlapping cluster patterns result in markedly lower performance, achieving only 67\% accuracy on SEED. Moreover, because HEDN depends on pre-established prototypes to align and classify target samples, its performance deteriorates when target data are extremely limited or entirely unseen, restricting its applicability in fully unknown or data-scarce settings.

Despite these limitations, HEDN remains effective with standard preprocessing, demonstrating reliable cross-subject emotion recognition by exploiting structured latent representations. Future research may explore constructing prototypes directly from raw EEG signals, improving resilience under sparse supervision, and enabling more flexible adaptation in real-world deployments where data availability and structure cannot be guaranteed.

\section{Conclusion}

In this work, we present HEDN, a novel reliability-aware Multi-Source Domain Adaptation framework designed to address the challenges of inter-subject variability and computational inefficiency in EEG emotion recognition. By introducing the Source Reliability Assessment mechanism, HEDN dynamically identifies the structural quality of each source domain and adaptively routes it into specialized learning pathways. The Easy Network employs prototype-based learning to utilize reliable domains for high-confidence pseudo-labeling, whereas the Hard Network leverages adversarial adaptation to enhance low-reliability domains. A cross-network consistency constraint further stabilizes learning by enforcing semantic coherence across heterogeneous supervision sources. Comprehensive experiments on SEED, SEED-IV, and DEAP datasets validate that HEDN achieves superior accuracy, robustness, and scalability compared with state-of-the-art approaches. Notably, the dual-branch design significantly reduces computational overhead, achieving faster adaptation with improved generalization performance, demonstrating the framework’s suitability for practical and resource-constrained EEG-based affective computing systems. Future work will focus on constructing prototypes directly from raw EEG signals to improve performance in data-scarce scenarios and further enhance resilience against extreme domain shifts.

\bibliographystyle{IEEEtran}
\bibliography{references}

@inproceedings{ali2016eeg,
  title={EEG-based emotion recognition approach for e-healthcare applications},
  author={Ali, Mouhannad and Mosa, Ahmad Haj and Al Machot, Fadi and Kyamakya, Kyandoghere},
  booktitle={2016 eighth international conference on ubiquitous and future networks (ICUFN)},
  pages={946--950},
  year={2016},
  organization={IEEE}
}

@article{brunner2015bnci,
  title={BNCI Horizon 2020: towards a roadmap for the BCI community},
  author={Brunner, Clemens and Birbaumer, Niels and Blankertz, Benjamin and Guger, Christoph and K{\"u}bler, Andrea and Mattia, Donatella and Mill{\'a}n, Jos{\'e} del R and Miralles, Felip and Nijholt, Anton and Opisso, Eloy and others},
  journal={Brain-computer interfaces},
  volume={2},
  number={1},
  pages={1--10},
  year={2015},
  publisher={Taylor \& Francis}
}

@article{li2022eeg,
  title={EEG based emotion recognition: A tutorial and review},
  author={Li, Xiang and Zhang, Yazhou and Tiwari, Prayag and Song, Dawei and Hu, Bin and Yang, Meihong and Zhao, Zhigang and Kumar, Neeraj and Marttinen, Pekka},
  journal={ACM Computing Surveys},
  volume={55},
  number={4},
  pages={1--57},
  year={2022},
  publisher={ACM New York, NY}
}

@article{zhong2020eeg,
  title={EEG-based emotion recognition using regularized graph neural networks},
  author={Zhong, Peixiang and Wang, Di and Miao, Chunyan},
  journal={IEEE Transactions on Affective Computing},
  volume={13},
  number={3},
  pages={1290--1301},
  year={2020},
  publisher={IEEE}
}

@article{li2019multisource,
  title={Multisource transfer learning for cross-subject EEG emotion recognition},
  author={Li, Jinpeng and Qiu, Shuang and Shen, Yuan-Yuan and Liu, Cheng-Lin and He, Huiguang},
  journal={IEEE transactions on cybernetics},
  volume={50},
  number={7},
  pages={3281--3293},
  year={2019},
  publisher={IEEE}
}

@inproceedings{wang2024dmmr,
  title={DMMR: Cross-subject domain generalization for EEG-based emotion recognition via denoising mixed mutual reconstruction},
  author={Wang, Yiming and Zhang, Bin and Tang, Yujiao},
  booktitle={Proceedings of the AAAI conference on artificial intelligence},
  volume={38},
  pages={628--636},
  year={2024}
}

@article{ganin2016domain,
  title={Domain-adversarial training of neural networks},
  author={Ganin, Yaroslav and Ustinova, Evgeniya and Ajakan, Hana and Germain, Pascal and Larochelle, Hugo and Laviolette, Fran{\c{c}}ois and March, Mario and Lempitsky, Victor},
  journal={Journal of machine learning research},
  volume={17},
  number={59},
  pages={1--35},
  year={2016}
}

@article{zheng2015investigating,
  title={Investigating critical frequency bands and channels for EEG-based emotion recognition with deep neural networks},
  author={Zheng, Wei-Long and Lu, Bao-Liang},
  journal={IEEE Transactions on autonomous mental development},
  volume={7},
  number={3},
  pages={162--175},
  year={2015},
  publisher={IEEE}
}

@article{zheng2018emotionmeter,
  title={Emotionmeter: A multimodal framework for recognizing human emotions},
  author={Zheng, Wei-Long and Liu, Wei and Lu, Yifei and Lu, Bao-Liang and Cichocki, Andrzej},
  journal={IEEE transactions on cybernetics},
  volume={49},
  number={3},
  pages={1110--1122},
  year={2018},
  publisher={IEEE}
}

@article{abdolzadegan2020robust,
  title={A robust method for early diagnosis of autism spectrum disorder from EEG signals based on feature selection and DBSCAN method},
  author={Abdolzadegan, Donya and Moattar, Mohammad Hossein and Ghoshuni, Majid},
  journal={Biocybernetics and Biomedical Engineering},
  volume={40},
  number={1},
  pages={482--493},
  year={2020},
  publisher={Elsevier}
}

@article{khosla2020supervised,
  title={Supervised contrastive learning},
  author={Khosla, Prannay and Teterwak, Piotr and Wang, Chen and Sarna, Aaron and Tian, Yonglong and Isola, Phillip and Maschinot, Aaron and Liu, Ce and Krishnan, Dilip},
  journal={Advances in neural information processing systems},
  volume={33},
  pages={18661--18673},
  year={2020}
}

@article{chen2021ms,
  title={MS-MDA: Multisource marginal distribution adaptation for cross-subject and cross-session EEG emotion recognition},
  author={Chen, Hao and Jin, Ming and Li, Zhunan and Fan, Cunhang and Li, Jinpeng and He, Huiguang},
  journal={Frontiers in Neuroscience},
  volume={15},
  pages={778488},
  year={2021},
  publisher={Frontiers Media SA}
}

@article{yang2024spectral,
  title={Spectral-spatial attention alignment for multi-source domain adaptation in EEG-based emotion recognition},
  author={Yang, Yi and Wang, Ze and Tao, Wei and Liu, Xucheng and Jia, Ziyu and Wang, Boyu and Wan, Feng},
  journal={IEEE Transactions on Affective Computing},
  volume={15},
  number={4},
  pages={2012--2024},
  year={2024},
  publisher={IEEE}
}

@article{li2024gusa,
  title={Gusa: Graph-based unsupervised subdomain adaptation for cross-subject EEG emotion recognition},
  author={Li, Xiaojun and Chen, CL Philip and Chen, Bianna and Zhang, Tong},
  journal={IEEE Transactions on Affective Computing},
  volume={15},
  number={3},
  pages={1451--1462},
  year={2024},
  publisher={IEEE}
}

@article{wang2024multi,
  title={Multi-source Selective Graph Domain Adaptation Network for cross-subject EEG emotion recognition},
  author={Wang, Jing and Ning, Xiaojun and Xu, Wei and Li, Yunze and Jia, Ziyu and Lin, Youfang},
  journal={Neural Networks},
  volume={180},
  pages={106742},
  year={2024},
  publisher={Elsevier}
}

@article{ma2023cross,
  title={Cross-subject emotion recognition based on domain similarity of EEG signal transfer learning},
  author={Ma, Yuliang and Zhao, Weicheng and Meng, Ming and Zhang, Qizhong and She, Qingshan and Zhang, Jianhai},
  journal={IEEE Transactions on Neural Systems and Rehabilitation Engineering},
  volume={31},
  pages={936--943},
  year={2023},
  publisher={IEEE}
}

@article{hong2024adaptive,
  title={Adaptive Domain Alignment Neural Networks for Cross-Domain EEG Emotion Recognition},
  author={Hong, Xuezhu and Du, Changde and He, Huiguang},
  journal={IEEE Transactions on Affective Computing},
  year={2024},
  publisher={IEEE}
}

@article{li2022dynamic,
  title={Dynamic domain adaptation for class-aware cross-subject and cross-session EEG emotion recognition},
  author={Li, Zhunan and Zhu, Enwei and Jin, Ming and Fan, Cunhang and He, Huiguang and Cai, Ting and Li, Jinpeng},
  journal={IEEE Journal of Biomedical and Health Informatics},
  volume={26},
  number={12},
  pages={5964--5973},
  year={2022},
  publisher={IEEE}
}

@article{guo2023multi,
  title={Multi-source domain adaptation with spatio-temporal feature extractor for EEG emotion recognition},
  author={Guo, Wenhui and Xu, Guixun and Wang, Yanjiang},
  journal={Biomedical Signal Processing and Control},
  volume={84},
  pages={104998},
  year={2023},
  publisher={Elsevier}
}

@article{chen2024gddn,
  title={GDDN: Graph domain disentanglement network for generalizable EEG emotion recognition},
  author={Chen, Bianna and Chen, CL Philip and Zhang, Tong},
  journal={IEEE Transactions on Affective Computing},
  volume={15},
  number={3},
  pages={1739--1753},
  year={2024},
  publisher={IEEE}
}

@article{maaten2008visualizing,
  title={Visualizing data using t-SNE},
  author={Maaten, Laurens van der and Hinton, Geoffrey},
  journal={Journal of machine learning research},
  volume={9},
  number={Nov},
  pages={2579--2605},
  year={2008}
}

@inproceedings{wood2024cluster,
  title={Cluster Triplet Loss for Unsupervised Domain Adaptation on Histology Images},
  author={Wood, Ruby and Domingo, Enric and Koelzer, Viktor Hendrik and Maughan, Timothy S and Rittscher, Jens},
  booktitle={Proceedings of the IEEE/CVF Conference on Computer Vision and Pattern Recognition},
  pages={5122--5131},
  year={2024}
}

@article{she2022multi,
  title={Multi-source manifold feature transfer learning with domain selection for brain-computer interfaces},
  author={She, Qingshan and Cai, Yinhao and Du, Shengzhi and Chen, Yun},
  journal={Neurocomputing},
  volume={514},
  pages={313--327},
  year={2022},
  publisher={Elsevier}
}

@inproceedings{zhao2020multi,
  title={Multi-source distilling domain adaptation},
  author={Zhao, Sicheng and Wang, Guangzhi and Zhang, Shanghang and Gu, Yang and Li, Yaxian and Song, Zhichao and Xu, Pengfei and Hu, Runbo and Chai, Hua and Keutzer, Kurt},
  booktitle={Proceedings of the AAAI conference on artificial intelligence},
  volume={34},
  pages={12975--12983},
  year={2020}
}

@article{liu2024capsnet,
  title={DA-CapsNet: A multi-branch capsule network based on adversarial domain adaption for cross-subject EEG emotion recognition},
  author={Liu, Shuaiqi and Wang, Zeyao and An, Yanling and Li, Bing and Wang, Xinrui and Zhang, Yudong},
  journal={Knowledge-Based Systems},
  volume={283},
  pages={111137},
  year={2024},
  publisher={Elsevier}
}

@article{wang2021deep,
  title={A deep multi-source adaptation transfer network for cross-subject electroencephalogram emotion recognition},
  author={Wang, Fei and Zhang, Weiwei and Xu, Zongfeng and Ping, Jingyu and Chu, Hao},
  journal={Neural Computing and Applications},
  volume={33},
  number={15},
  pages={9061--9073},
  year={2021},
  publisher={Springer}
}

@article{shen2022contrastive,
  title={Contrastive learning of subject-invariant EEG representations for cross-subject emotion recognition},
  author={Shen, Xinke and Liu, Xianggen and Hu, Xin and Zhang, Dan and Song, Sen},
  journal={IEEE Transactions on Affective Computing},
  volume={14},
  number={3},
  pages={2496--2511},
  year={2022},
  publisher={IEEE}
}

@article{zhou2023pr,
  title={PR-PL: A novel prototypical representation based pairwise learning framework for emotion recognition using EEG signals},
  author={Zhou, Rushuang and Zhang, Zhiguo and Fu, Hong and Zhang, Li and Li, Linling and Huang, Gan and Li, Fali and Yang, Xin and Dong, Yining and Zhang, Yuan-Ting and others},
  journal={IEEE Transactions on Affective Computing},
  volume={15},
  number={2},
  pages={657--670},
  year={2023},
  publisher={IEEE}
}

@article{wang2025fine,
  title={Fine-Grained Label Propagation via Density-Based Prototype Matching for Cross-Subject EEG Emotion Recognition},
  author={Wang, Qiang and Yang, Liying and Zhang, Qian and Du, Jingtao and Ye, Yumeng},
  journal={Knowledge-Based Systems},
  pages={114650},
  year={2025},
  publisher={Elsevier}
}

@article{koelstra2011deap,
  title={Deap: A database for emotion analysis; using physiological signals},
  author={Koelstra, Sander and Muhl, Christian and Soleymani, Mohammad and Lee, Jong-Seok and Yazdani, Ashkan and Ebrahimi, Touradj and Pun, Thierry and Nijholt, Anton and Patras, Ioannis},
  journal={IEEE transactions on affective computing},
  volume={3},
  number={1},
  pages={18--31},
  year={2011},
  publisher={IEEE}
}

@ARTICLE{huang2025cda,
  author={Huang, He and Si, Xiaopeng and Han, Yumeng and Ming, Dong},
  journal={IEEE Transactions on Affective Computing}, 
  title={A Novel Conditional Adversarial Domain Adaptation Network for EEG Cross-subject Emotion Recognition}, 
  year={2025},
  volume={},
  number={},
  pages={1-13},
  doi={10.1109/TAFFC.2025.3588873}}

@inproceedings{li2018cross,
  title={Cross-subject emotion recognition using deep adaptation networks},
  author={Li, He and Jin, Yi-Ming and Zheng, Wei-Long and Lu, Bao-Liang},
  booktitle={International conference on neural information processing},
  pages={403--413},
  year={2018},
  organization={Springer}
}

@article{cheng2025disd,
  title={DISD-Net: A dynamic interactive network with self-distillation for cross-subject multi-modal emotion recognition},
  author={Cheng, Cheng and Liu, Wenzhe and Wang, Xinying and Feng, Lin and Jia, Ziyu},
  journal={IEEE Transactions on Multimedia},
  year={2025},
  publisher={IEEE}
}

@article{du2020efficient,
  title={An efficient LSTM network for emotion recognition from multichannel EEG signals},
  author={Du, Xiaobing and Ma, Cuixia and Zhang, Guanhua and Li, Jinyao and Lai, Yu-Kun and Zhao, Guozhen and Deng, Xiaoming and Liu, Yong-Jin and Wang, Hongan},
  journal={IEEE Transactions on Affective Computing},
  volume={13},
  number={3},
  pages={1528--1540},
  year={2020},
  publisher={IEEE}
}

@article{liu2024enhancing,
  title={Enhancing EEG-Based Cross-Subject Emotion Recognition via Adaptive Source Joint Domain Adaptation},
  author={Liu, Ke and Luo, Xin and Zhu, Wenrui and Yu, Zhuliang and Yu, Hong and Xiao, Bin and Wu, Wei},
  journal={IEEE Transactions on Affective Computing},
  year={2024},
  publisher={IEEE}
}

@article{zhou2025enhancing,
  title={Enhancing cross-dataset eeg emotion recognition: A novel approach with emotional eeg style transfer network},
  author={Zhou, Yijin and Li, Fu and Li, Yang and Ji, Youshuo and Zhang, Lijian and Chen, Yuanfang and Wang, Huaning},
  journal={IEEE Transactions on Affective Computing},
  year={2025},
  publisher={IEEE}
}

@article{lan2018domain,
  title={Domain adaptation techniques for EEG-based emotion recognition: a comparative study on two public datasets},
  author={Lan, Zirui and Sourina, Olga and Wang, Lipo and Scherer, Reinhold and M{\"u}ller-Putz, Gernot R},
  journal={IEEE Transactions on Cognitive and Developmental Systems},
  volume={11},
  number={1},
  pages={85--94},
  year={2018},
  publisher={IEEE}
}

@article{li2022gmss,
  title={GMSS: Graph-based multi-task self-supervised learning for EEG emotion recognition},
  author={Li, Yang and Chen, Ji and Li, Fu and Fu, Boxun and Wu, Hao and Ji, Youshuo and Zhou, Yijin and Niu, Yi and Shi, Guangming and Zheng, Wenming},
  journal={IEEE Transactions on Affective Computing},
  volume={14},
  pages={2512--2525},
  year={2022},
  publisher={IEEE}
}

@inproceedings{song2020instance,
  title={Instance-adaptive graph for EEG emotion recognition},
  author={Song, Tengfei and Liu, Suyuan and Zheng, Wenming and Zong, Yuan and Cui, Zhen},
  booktitle={Proceedings of the AAAI Conference on Artificial Intelligence},
  volume={34},
  pages={2701--2708},
  year={2020}
}

@article{snell2017prototypical,
  title={Prototypical networks for few-shot learning},
  author={Snell, Jake and Swersky, Kevin and Zemel, Richard},
  journal={Advances in neural information processing systems},
  volume={30},
  year={2017}
}

@article{ji2020improved,
  title={Improved prototypical networks for few-shot learning},
  author={Ji, Zhong and Chai, Xingliang and Yu, Yunlong and Pang, Yanwei and Zhang, Zhongfei},
  journal={Pattern Recognition Letters},
  volume={140},
  pages={81--87},
  year={2020},
  publisher={Elsevier}
}

@article{schubert2017dbscan,
  title={DBSCAN revisited, revisited: why and how you should (still) use DBSCAN},
  author={Schubert, Erich and Sander, J{\"o}rg and Ester, Martin and Kriegel, Hans Peter and Xu, Xiaowei},
  journal={ACM Transactions on Database Systems (TODS)},
  volume={42},
  number={3},
  pages={1--21},
  year={2017},
  publisher={Acm New York, NY, USA}
}

@article{han2018co,
  title={Co-teaching: Robust training of deep neural networks with extremely noisy labels},
  author={Han, Bo and Yao, Quanming and Yu, Xingrui and Niu, Gang and Xu, Miao and Hu, Weihua and Tsang, Ivor and Sugiyama, Masashi},
  journal={Advances in neural information processing systems},
  volume={31},
  year={2018}
}

@inproceedings{hu2023swl,
  title={Swl-adapt: An unsupervised domain adaptation model with sample weight learning for cross-user wearable human activity recognition},
  author={Hu, Rong and Chen, Ling and Miao, Shenghuan and Tang, Xing},
  booktitle={Proceedings of the AAAI Conference on artificial intelligence},
  volume={37},
  number={5},
  pages={6012--6020},
  year={2023}
}

@article{ahn2016exploring,
  title={Exploring neuro-physiological correlates of drivers' mental fatigue caused by sleep deprivation using simultaneous EEG, ECG, and fNIRS data},
  author={Ahn, Sangtae and Nguyen, Thien and Jang, Hyojung and Kim, Jae G and Jun, Sung C},
  journal={Frontiers in human neuroscience},
  volume={10},
  pages={219},
  year={2016},
  publisher={Frontiers Media SA}
}

@article{mehmood2023deep,
  title={Deep learning-based construction equipment operators’ mental fatigue classification using wearable EEG sensor data},
  author={Mehmood, Imran and Li, Heng and Qarout, Yazan and Umer, Waleed and Anwer, Shahnawaz and Wu, Haitao and Hussain, Mudasir and Antwi-Afari, Maxwell Fordjour},
  journal={Advanced Engineering Informatics},
  volume={56},
  pages={101978},
  year={2023},
  publisher={Elsevier}
}

@article{zhou2024eegmatch,
  title={Eegmatch: Learning with incomplete labels for semisupervised eeg-based cross-subject emotion recognition},
  author={Zhou, Rushuang and Ye, Weishan and Zhang, Zhiguo and Luo, Yanyang and Zhang, Li and Li, Linling and Huang, Gan and Dong, Yining and Zhang, Yuan-Ting and Liang, Zhen},
  journal={IEEE Transactions on Neural Networks and Learning Systems},
  year={2024},
  publisher={IEEE}
}

\end{document}